\documentclass[11pt]{article}
\usepackage[a4paper, top = 1.25in, bottom = 1.5in]{geometry}
\usepackage{package_perso} 
\usepackage{tocloft}
\begin{document}
\begin{titlepage}

\newcommand{\HRule}{\rule{\linewidth}{0.5mm}} 
\begin{center}
\LARGE{ DTM-based Filtrations }
\vspace{1cm}

\large{
Hirokazu \textsc{Anai}$^\dagger$, 
Frédéric \textsc{Chazal}$^*$, 
Marc \textsc{Glisse}$^*$, 
Yuichi \textsc{Ike}$^\dagger$, \linebreak
Hiroya \textsc{Inakoshi}$^\dagger$, 
Raphaël \textsc{Tinarrage}$^{*\star}$, 
Yuhei \textsc{Umeda}$^\dagger$ 
}

\vspace{.5cm} 
\large{ 
*Datashape, Inria Paris-Saclay \linebreak
$\dagger$ Fujitsu Laboratories, AI Lab \linebreak
$\star$ LMO, Université Paris-Saclay
}
%\vspace{.8cm} 
%\today
\end{center}
\vspace{.2cm} 

\paragraph{Abstract.} Despite strong stability properties, the persistent homology of filtrations classically used in Topological Data Analysis, such as, e.g. the \v{C}ech or Vietoris-Rips filtrations, are very sensitive to the presence of outliers in the data from which they are computed. 
In this paper, we introduce and study a new family of filtrations, the DTM-filtrations, 
built on top of point clouds in the Euclidean space which are more robust to noise and outliers. 
The approach adopted in this work relies on the notion of distance-to-measure functions, and extends some previous work on the approximation of such functions. 

\paragraph{Numerical experiments.} 
A Python notebook here \url{https://github.com/GUDHI/TDA-tutorial/blob/master/Tuto-GUDHI-DTM-filtrations.ipynb}
and some animations \url{https://www.youtube.com/playlist?list=PL_FkltNTtklAOt4dkygG8Zv_-wUvlaz07}.

\vspace{.2cm} 
\tableofcontents
\end{titlepage}

\newpage
\addtocounter{page}{1}
\section{Introduction}
The inference of relevant topological properties of data represented as point clouds in Euclidean spaces is a central challenge in Topological Data Analysis (TDA).   

Given a (finite) set of points $X$ in $\R^d$, persistent homology provides a now classical and powerful tool to construct persistence diagrams whose points can be interpreted as homological features of $X$ at different scales. 
These persistence diagrams are obtained from {\em filtrations}, i.e. nested families of subspaces or simplicial complexes, built on top of $X$. 
Among the many filtrations available to the user, unions of growing balls $\cup_{x \in X} \overline B(x,t)$ (sublevel sets of distance functions), $t \in \R^+$, and their nerves, the \v{C}ech complex filtration, or its usually easier to compute variation, the Vietoris-Rips filtration, are widely used. The main theoretical advantage of these filtrations is that they have been shown to produce persistence diagrams that are stable with respect to perturbations of $X$ in the Hausdorff metric \cite{Chazal_Geometriccomplexes}.

Unfortunately, the Hausdorff distance turns out to be very sensitive to noise and outliers, preventing the direct use of distance functions and classical \v{C}ech or Vietoris-Rips filtrations to infer relevant topological properties from real noisy data. 
Several attempts have been made in the recent years to overcome this issue. 
Among them, the filtration defined by the sublevel sets of the distance-to-measure (DTM) function introduced in \cite{Chazal_Geometricinference}, and some of its variants \cite{pwz-gikde-15}, have been proven to provide relevant information about the geometric structure underlying the data. Unfortunately, from a practical perspective, the exact computation of the sublevel sets filtration of the DTM, that boils down to the computation of a $k$-th order Vorono\"i diagram, and its persistent homology turn out to be far too expensive in most cases. 
To address this problem, \cite{guibas2013witnessed} introduces a variant of the DTM function, the witnessed $k$-distance, whose persistence is easier to compute and proves that the witnessed $k$-distance approximates the DTM persistence up to a fixed additive constant. In \cite{Buchet_Efficient, Buchet_Thesis}, a weighted version of the Vietoris-Rips complex filtration is introduced to approximate the persistence of the DTM function, and several stability and approximation results, comparable to the ones of \cite{guibas2013witnessed}, are established. Another kind of weighted Vietoris-Rips complex is presented in \cite{bell2017weighted}.

\paragraph{Contributions.} In this paper, we introduce and study a new family of filtrations based on the notion of DTM. Our contributions are the following:   
\begin{itemize}
\item Given a set $X \subset \R^d$, a weight function $f$ defined on $X$ and $p \in [1, +\infty]$, we introduce the weighted \v{C}ech and Rips filtrations that extend the notion of sublevel set filtration of power distances of \cite{Buchet_Efficient}. Using classical results, we show that these filtrations are stable with respect to perturbations of $X$ in the Hausdorff metric and perturbations of $f$ with respect to the sup norm (Propositions \ref{proposition_stability_1} and \ref{proposition_stability_2}).
\item For a general function $f$, the stability results of the weighted \v{C}ech and Rips filtrations are not suited to deal with noisy data or data containing outliers. We consider the case where $f$ is the empirical DTM-function associated to the input point cloud. In this case, we show an outliers-robust stability result: given two point clouds $X, Y \subseteq \R^d$, the closeness between the persistence diagrams of the resulting filtrations relies on the existence of a subset of $X$ which is both close to $X$ and $Y$ in the Wasserstein metric (Theorems \ref{prop_stability_p=1} and \ref{thm:DTM-main}). 
\end{itemize}
%---here the point clouds are seen as empirical measures---

\begin{figure}[H]
\centering
\begin{center}
\begin{tabular}{ | m{2.5cm} | m{5.2cm}| m{5.2cm} | } 
 \hline
 & Time series without rapid shift & Time series with rapid shift \\
 \hline
 Time series and time delay embedding 
 & \vspace{0.1cm}\includegraphics[width=.45\linewidth]{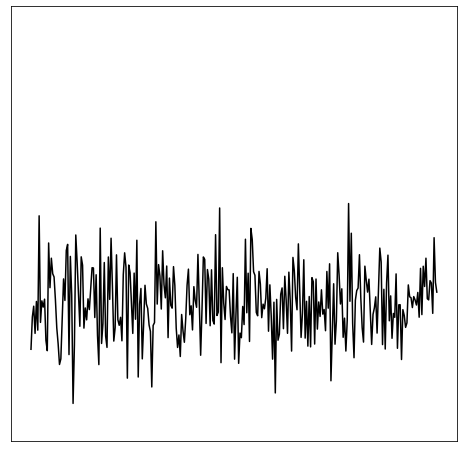}
 \includegraphics[width=.50\linewidth]{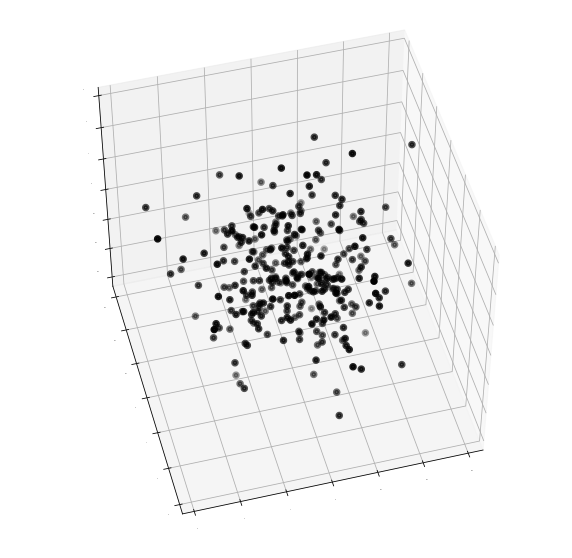}  
 & \vspace{0.1cm}\includegraphics[width=.45\linewidth]{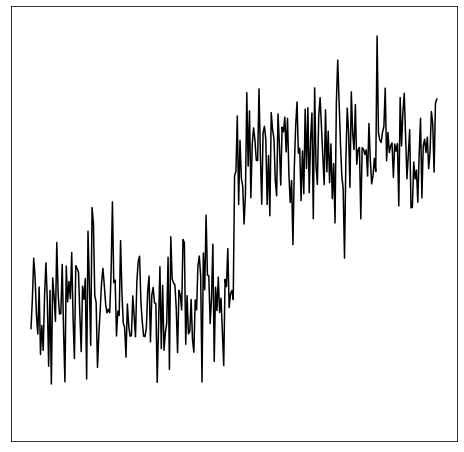}  
\includegraphics[width=.50\linewidth]{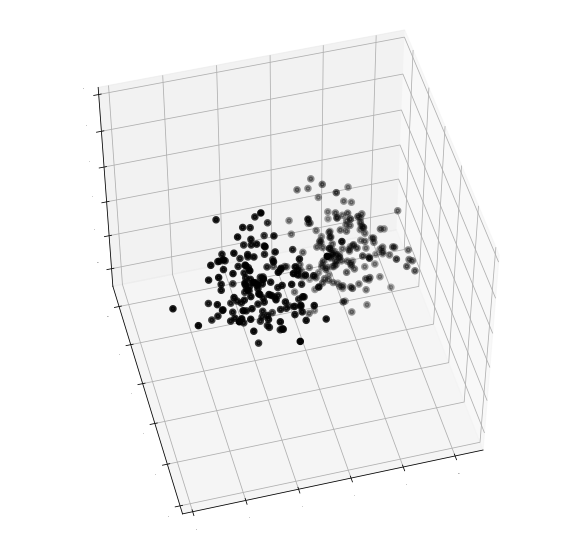} \\
\hline
  Conventional filtration
 & \vspace{0.1cm}~~~~~~~~\includegraphics[width=.50\linewidth]{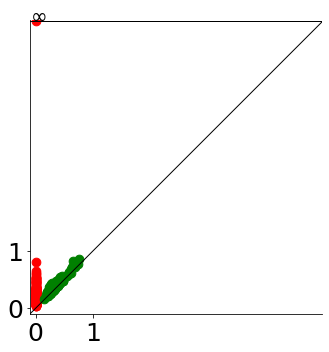}  
 & \vspace{0.1cm}~~~~~~~~\includegraphics[width=.50\linewidth]{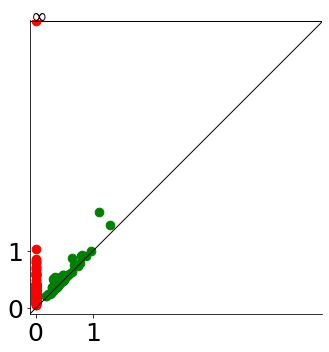}  \\
\hline
 DTM-filtration 
 &  \vspace{0.1cm}~~~~~~~~\includegraphics[width=.50\linewidth]{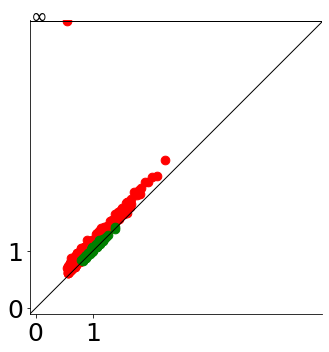}  
 & \vspace{0.1cm}~~~~~~~~\includegraphics[width=.50\linewidth]{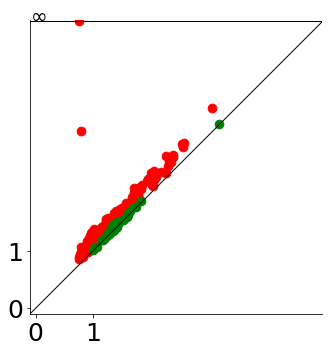}  \\
\hline
\end{tabular}
\end{center}
%\begin{tabularx}{0.95\linewidth} { 
%  | >{\raggedright\arraybackslash}X 
%  | >{\centering\arraybackslash}X 
%  | >{\centering\arraybackslash}X | }
% \hline
% & Time series without rapid shift & Time series with rapid shift \\
% \hline
% Time series and time delay embedding  \vspace{0.2cm}  
% & \includegraphics[width=.45\linewidth]{Intro_1.png}
% \includegraphics[width=.45\linewidth]{Intro_2.png}  
% & \includegraphics[width=.45\linewidth]{Intro_5.png}  
%\includegraphics[width=.45\linewidth]{Intro_6.png} \\
%\hline
% ~ \vspace{1cm} Conventional filtration \vspace{1cm} 
% & \includegraphics[width=.85\linewidth]{Intro_3.png}  
% & \includegraphics[width=.95\linewidth]{Intro_7.png}  \\
%\hline
% DTM-filtration  \vspace{0.2cm} 
% & \includegraphics[width=.95\linewidth]{Intro_4.png}  
% & \includegraphics[width=.95\linewidth]{Intro_8.png}  \\
%\hline
%\end{tabularx}
\caption{A synthetic example comparing Vietoris-Rips filtration to DTM filtration. The first row represents two time series with very different behavior and their embedding into $\R^3$ (here a series $(x_1, x_2,\dots, x_{n})$ is converted in the 3D point cloud 
$\{ (x_1,x_2,x_3), (x_2,x_3,x_4), \dots, (x_{n-2},x_{n-1},x_{n}) \}$). The second row shows the persistence diagrams of the Vietoris-Rips filtration built on top of the two point clouds (red and green points represent respectively the $0$-dimensional $1$-dimensional diagrams); one observes that the diagrams do not clearly `detect' the different behavior of the time series. The third row shows the persistence diagrams of the DTM filtration built on top  of the two point clouds; a red point clearly appears away from the diagonal in the second diagram that highlights the rapid shift occurring in the second time series.}   
\label{fig:ExampleIntro}
\end{figure}

\paragraph{Practical motivations.}  Even though this aspect is not considered in this paper, it is interesting to mention that the DTM filtration was first experimented in the setting of an industrial research project whose goal was to address an anomaly detection problem from inertial sensor data in bridge and building monitoring \cite{fujitsumarch2018}. In this problem, the input data comes as time series measuring the acceleration of devices attached to the monitored bridge/building. Using sliding windows and time-delay embedding, these times series are converted into a series of fixed size point clouds in $\R^d$. Filtrations are then built on top of these point clouds and their persistence is computed, giving rise to a time-dependent sequence of persistence diagrams that are then used to detect anomalies or specific features occurring along the time \cite{seversky2016time,umeda2017time}. In this practical setting it turned out that the DTM filtrations reveal to be not only more resilient to noise but also able to better highlight topological features in the data than the standard Vietoris-Rips filtrations, as illustrated on a basic synthetic example on Figure \ref{fig:ExampleIntro}. One of the goals of the present work is to provide theoretical foundations to these promising experimental results by studying the stability properties of the DTM filtrations. 

\paragraph{Organisation of the paper.} Preliminary definitions, notations, and basic notions on filtrations and persistence modules are recalled in Section \ref{sec:preliminary}. The weighted \v{C}ech and Vietoris-Rips filtrations are introduced in Section \ref{section_Vp_modules}, where their stability properties are established. The DTM-filtrations are introduced in Section \ref{section_DTM}. Their main stability properties are established in Theorems \ref{prop_stability_p=1} and \ref{thm:DTM-main}, and their relation with the sublevel set filtration of the DTM-functions is established in Proposition \ref{prop:DTMvsDTMfilt}. For the clarity of the paper, the proofs of several lemmas have been postponed to the appendices. 

The various illustrations and experiments of this paper have been computed with the GUDHI library on Python \cite{gudhilib}. 

\paragraph{Acknowledgements.} This work was partially supported by a collaborative research agreement between Inria and Fujitsu, and the Advanced Grant of the European Research Council GUDHI (Geometric Understanding in Higher Dimensions).

\section{Filtrations and interleaving distance} \label{sec:preliminary}
In this subsection, we consider interleavings of filtrations, interleavings of persistence modules and their associated pseudo-distances. Their definitions, restricted to the setting of the paper, are briefly recalled in this section.  

Let $T = \R^+$ and $E = \R^d$ endowed with the standard Euclidean norm.  

\paragraph{Filtrations of sets and simplicial complexes.}
A family of subsets $(V^t)_{t \in T}$ of $E = \R^d$ is a {\em filtration} if it is non-decreasing for the inclusion, i.e. for any $s, t \in T$, if $s \leq t$ then $V^s \subseteq V^t$. 
Given $\epsilon\geq 0$, two filtrations $(V^t)_{t \in T}$ and $(W^t)_{t \in T}$ of $E$ are {\em $\epsilon$-interleaved} if, for every $t \in T$, $V^t \subseteq W^{t+\epsilon}$ and $W^t \subseteq V^{t+\epsilon}$. 
The interleaving pseudo-distance between $(V^t)_{t \in T}$ and $(W^t)_{t \in T}$ is defined as the infimum of such $\epsilon$:
$$d_i((V^t)_{t \in T}, (W^t)_{t \in T}) = \inf \{ \epsilon: (V^t) \ \mbox{\rm and} \ (W^t) \ \mbox{\rm are $\epsilon$-interleaved} \}.$$

Filtrations of simplicial complexes and their interleaving distance are similarly defined: 
given a set $X$ and an abstract simplex $S$ with vertex set $X$, a {\em filtration of $S$} is a non-decreasing family $(S^t)_{t \in T}$ of subcomplexes of $S$. The interleaving pseudo-distance between two filtrations $(S_1^t)_{t \in T}$ and $(S_2^t)_{t \in T}$ of $S$ is the infimum of the $\epsilon \geq 0$ such that they are $\epsilon$-interleaved, i.e. for any $t \in T$, $S_1^{t} \subseteq S_2^{t+\epsilon}$ and  $S_2^{t} \subseteq S_1^{t+\epsilon}$.

Notice that the interleaving distance is only a pseudo-distance, as two distinct filtrations may have zero interleaving distance.  

\paragraph{Persistence modules.}
Let $k$ be a field. A {\em persistence module} $\V$ over $T = \R^+$ is a pair $\V = ((\V^t)_{t\in T}, (v_s^t)_{s\leq t \in T})$ where $(\V^t)_{t\in T}$ is a family of $k$-vector spaces, and $(v_s^t\colon \V^s \rightarrow \V^t)_{s\leq t \in T}$ a family of linear maps such that:
\begin{itemize}
\item for every $t\in T$, $v_t^t\colon V^t \rightarrow V^t$ is the identity map,
\item for every $r, s,t\in T$ such that $r\leq s\leq t$, $v_s^t \circ v_r^s = v_r^t$.
\end{itemize}  
Given $\epsilon \geq 0$, an {\em $\epsilon$-morphism} between two persistence modules $\V$ and $\W$ is a family of linear maps $(\phi_t\colon \V^t \rightarrow \W^{t+\epsilon})_{t \in T}$ such that the following diagrams commute for every $s \leq t \in T$:
\begin{center}
\begin{tikzcd}
\V^s \arrow["\phi_s", d] \arrow[r, "v_s^t"] & \arrow["\phi_t", d] \V^t \\
\W^{s+\epsilon} \arrow[r, "w_{s+\epsilon}^{t+\epsilon}"] & \W^{t+\epsilon}
\end{tikzcd}
\end{center}
If $\epsilon = 0$ and each $\phi_t$ is an isomorphism, the family $(\phi_t)_{t \in T}$ is said to be an {\em isomorphism} of persistence modules.

An {\em $\epsilon$-interleaving} between two persistence modules $\V$ and $\W$ is a pair of $\epsilon$-morphisms $(\phi_t\colon \V^t \rightarrow \W^{t+\epsilon})_{t \in T}$ and $(\psi_t\colon \W^t \rightarrow \V^{t+\epsilon})_{t \in T}$ such that the following diagrams commute for every $t \in T$: 
\begin{center}
\begin{minipage}[t]{0.4\textwidth}
\centering
\begin{tikzcd}
\V^t \arrow[dr, "\phi_t"] \arrow[rr, "v_t^{t+2\epsilon}"] & & \V^{t+2\epsilon} \\
& \W^{t+\epsilon} \arrow[ur, "\psi_{t+\epsilon}"] & 
\end{tikzcd}
\end{minipage}
\begin{minipage}[t]{0.4\textwidth}
\centering
\begin{tikzcd}
& \V^{t+\epsilon} \arrow[dr, "\phi_{t+\epsilon}"] & \\
\W^t \arrow[ur, "\psi_t"] \arrow[rr, "w_t^{t+2\epsilon}"] & & \W^{t+2\epsilon} 
\end{tikzcd}
\end{minipage}
\end{center}
The interleaving pseudo-distance between $\V$ and $\W$ is defined as 
$$d_i(\V, \W) = \inf \{\epsilon \geq 0, \V \text{ and } \W \text{ are } \epsilon \text{-interleaved}\}.$$

In some cases, the proximity between persistence modules is expressed with a function. Let $\eta\colon T \rightarrow T$ be a non-decreasing function such that for any $t \in T$, $\eta(t) \geq t$. A $\eta$-interleaving between two persistence modules $\V$ and $\W$ is a pair of families of linear maps $(\phi_t\colon \V^t \rightarrow \W^{\eta(t) })_{t \in T}$ and $(\psi_t\colon \W^t \rightarrow \V^{\eta(t)})_{t \in T}$ such that the following diagrams commute for every $t \in T$: 
\begin{center}
\begin{minipage}[t]{0.4\textwidth}
\centering
\begin{tikzcd}
\V^t \arrow[dr, "\phi_t"] \arrow[rr, "v_t^{\eta(\eta(t))}"] & & \V^{\eta(\eta(t))} \\
& \W^{\eta(t)} \arrow[ur, "\psi_{\eta(t)}"] & 
\end{tikzcd}
\end{minipage}
\begin{minipage}[t]{0.4\textwidth}
\centering
\begin{tikzcd}
& \V^{\eta(t)} \arrow[dr, "\phi_{\eta(t)}"] & \\
\W^t \arrow[ur, "\psi_t"] \arrow[rr, "v_t^{\eta(\eta(t))}"] & & \W^{\eta(\eta(t))} 
\end{tikzcd}
\end{minipage}
\end{center}
When $\eta$ is $t \mapsto t +c$ for some $c\geq 0$, it is called an additive $c$-interleaving and corresponds with the previous definition. When $\eta$ is $t \mapsto ct$ for some $c \geq 1$, it is called a multiplicative $c$-interleaving. 

A persistence module $\V$ is said to be {\em $q$-tame} if for every $s,t \in T$ such that $s < t$, the map $v_s^t$ is of finite rank. The $q$-tameness of a persistence module ensures that we can define a notion of persistence diagram---see \cite{Chazal_Persistencemodules}. Moreover, given two $q$-tame persistence modules $\V, \W$ with persistence diagrams $D(\V), D(\W)$, the so-called isometry theorem states that $d_b(D(\V), D(\W)) = d_i(\V, \W)$ (\cite[Theorem~4.11]{Chazal_Persistencemodules}) where $d_b(\cdot,\cdot)$ denotes the bottleneck distance between diagrams. 

\paragraph{Relation between filtrations and persistence modules.}
Applying the homology functor to a filtration gives rise to a persistence module where the linear maps between homology groups are induced by the inclusion maps between sets (or simplicial complexes). As a consequence, if two filtrations are $\epsilon$-interleaved then their associated homology persistence modules are also $\epsilon$-interleaved, the interleaving homomorphisms being induced by the interleaving inclusion maps. Moreover, if the modules are $q$-tame, then the bottleneck distance between their persistence diagrams is upperbounded by $\epsilon$.

The filtrations considered in this paper are obtained as union of growing balls. Their associated persistence module is the same as the persistence module of a filtered simplicial complex via the persistent nerve lemma (\cite{Chazal_Towards}, Lemma 3.4). 
Indeed, consider a filtration $(V^t)_{t \in T}$ of $E$ and assume that there exists a family of points $(x_i)_{i \in I} \in E^I$ and a family of non-decreasing functions $r_i\colon T \rightarrow \R^+ \cup \{-\infty\}$, $i \in I$, such that, for every $t\in T$, $V^t$ is equal to the union of closed balls $\bigcup_I \overline B(x_i, r_i(t))$, with the convention $\overline B(x_i, -\infty) = \emptyset$. For every $t \in T$, let $\mathcal{V}^t$ denote the cover $\{\overline B(x_i, r_i(t)), i \in I\}$ of $V^t$, and $S^t$ be its nerve. Let $\V$ be the persistence module associated with the filtration $(V^t)_{t \in T}$, and $\V_\mathcal{N}$ the one associated with the simplicial filtration $(S^t)_{t \in T}$. Then $\V$ and $\V_\mathcal{N}$ are isomorphic persistence modules. In particular, if $\V$ is $q$-tame, $\V$ and $\V_\mathcal{N}$ have the same persistence diagrams. 

\section{Weighted \texorpdfstring{\v{C}ech}{Cech} filtrations}
\label{section_Vp_modules}

In order to define the DTM-filtrations, we go through an intermediate and more general construction, namely the weighted \v{C}ech filtrations. It generalizes the usual notion of \v{C}ech filtration of a subset of $\R^d$, and shares comparable regularity properties.

\subsection{Definition}
\begin{figure}[H]
\centering
\includegraphics[width=.8\linewidth]{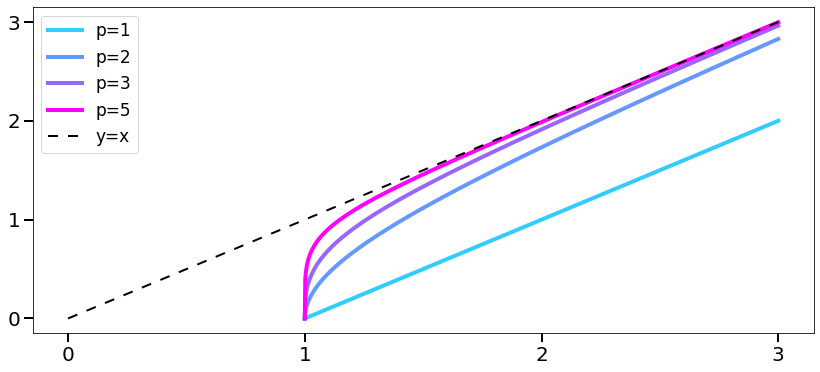}
\caption{Graph of $t \mapsto r_x(t)$ for $f(x) = 1$ and several values of $p$.}
\label{Fig_rx}
\end{figure}

In the rest of the paper, the Euclidean space $E = \R^d$, the index set $T=\R^+$ and a real number $p \geq 1$ are fixed.
Consider $X \subseteq E$ and $f\colon X \rightarrow \R^+$. 
For every $x \in X$ and $t \in T$, we define 
\[r_x(t) = 
\begin{cases}
	-\infty &\text{ if } t<f(x),\\    
	\big(t^p-f(x)^p\big)^\frac{1}{p} &\text{ otherwise.}\
\end{cases}\]
We denote by $\overline B_f(x,t)= \overline B(x, r_x(t))$ the closed Euclidean ball of center $x$ and radius $r_x(t)$.
By convention, a Euclidean ball of radius $-\infty$ is the empty set. 
For $p = \infty$, we also define
\[r_x(t) = 
\begin{cases}
	-\infty &\text{ if } t<f(x),\\    
	t &\text{ otherwise, }\
\end{cases}\]
and the balls $\overline B_f(x,t)= \overline B(x, r_x(t))$.
Some of these radius functions are represented in Figure \ref{Fig_rx}.

%When $p\in [1, +\infty[$, notice the following facts about $r_x(t)$ :
%\begin{itemize}
%\item it is twice differentiable and concave on $]f(x), +\infty[$, with 
%\[r_x'(t) = \frac{t^{p-1}}{\big( t^p-f(x)^p \big)^{1-\frac{1}{p}}} ~~~\text{and}~~~ r_x''(t) = -(p-1) f(x)^p \frac{t^{p-2}}{\big( t^p-f(x)^p %\big)^{2-\frac{1}{p}}}.\] 
%\item if $p>1$ and $f(x)\neq 0$, $\lim_{t \rightarrow f(x)} r_x'(t) = +\infty$.
%\item the following asymptotic expansion holds :
%\[r_x(t) = t -\frac{f(x)^p}{p}\frac{1}{t^{p-1}} + \underset{t\to \infty}{o}\big(\frac{1}{t^{p-1}}\big) \]
%\item for every $y \in E$, $y \in \overline B_f(x,t)$ is equivalent to $\big( \|x-y\|^p+f(x)^p \big)^\frac{1}{p} \leq t$
%\end{itemize}

\begin{definition}
Let $X \subseteq E$ and $f\colon X \rightarrow \R^+$. 
For every $t \in T$, we define the following set:
\[\CechFt{X,f}{t} = \bigcup_{x \in X} \overline{B}_f(x, t).\]
The family $\CechF{X,f} = (\CechFt{X,f}{t})_{t\geq 0}$ is a filtration of $E$.
It is called the \emph{weighted \v{C}ech filtration with parameters $(X,f,p)$}. 
We denote by $\CechFpers{X,f}$ its persistence (singular) homology module. 
\label{def_weightedCech}
\end{definition}

Note that $\CechF{X,f}$ and $\CechFpers{X,f}$ depend on fixed parameter $p$, that is not made explicit in the notation.
\medbreak

Introduce $\CechFcovert{X,f}{t} = \{\overline{B}_f(x, t)\}_{x \in X}$. It is a cover of $\CechFt{X,f}{t}$ by closed Euclidean balls. Let $\CechFnervet{X,f}{t}$ be the nerve of the cover $\CechFcovert{X,f}{t}$. It is a simplicial complex over the vertex set $X$. 
The family $\CechFnerve{X,f} = (\CechFnervet{X,f}{t})_{t\geq 0}$ is a filtered simplicial complex. We denote by $\CechFnervepers{X,f}$ its persistence (simplicial) homology module. As a consequence of the persistent nerve theorem \cite[Lemma 3.4]{Chazal_Towards}, $\CechFpers{X,f}$ and $\CechFnervepers{X,f}$ are isomorphic persistence modules.
\medbreak

When $f=0$, $\CechF{X,f}$ does not depend on $p \geq 1$, and it is the filtration of $E$ by the sublevel sets of the distance function to $X$. We denote it by $\CechF{X,0}$. The corresponding filtered simplicial complex, $\CechFnerve{X,0}$, is known as the usual \v{C}ech complex of $X$.

When $p=2$, the filtration value of $y \in E$, i.e. the infimum of the $t$ such that $y \in \CechFt{X,f}{t}$, is called the power distance of $y$ associated to the weighted set $(X, f)$ in \cite[Definition 4.1]{Buchet_Efficient}. The filtration $\CechF{X,f}$ is called the weighted \v{C}ech filtration (\cite[Definition 5.1]{Buchet_Efficient}).

\begin{figure}[H]
%   \centering
\hspace{-.2cm}
\begin{tabular}{cccc}
\includegraphics[width=3.2cm]{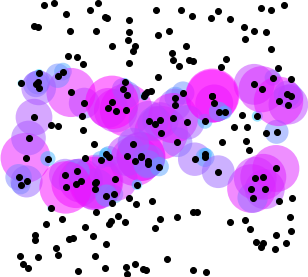}&
\includegraphics[width=3.2cm]{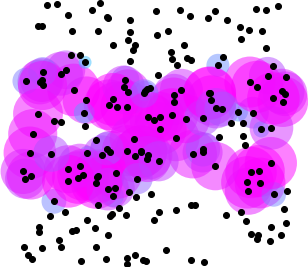}&
\includegraphics[width=3.2cm]{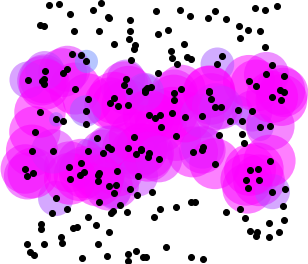}&
\includegraphics[width=3.2cm]{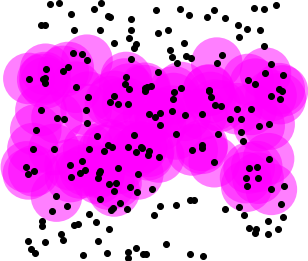}\\
 $p=1$ & $p=2$ & $p=3$ & $p=\infty$\\
\end{tabular}
\caption{The sets $\CechFt{X,f}{t}$ for $t=0{,}2$ and several values of $p$.}
\label{Fig_weightedCech} 
\end{figure}

\vspace{5mm} \noindent
\begin{minipage}[t]{0.69\textwidth}
\textbf{Example.} Consider the point cloud $X$ drawn on the right (black). It is a 200-sample of the uniform distribution on $[-1,1]^2 \subseteq \R^2$. We choose $f$ to be the distance function to the lemniscate of Bernoulli (magenta). 
Let $t = 0{,}2$. Figure \ref{Fig_weightedCech} represents the sets $\CechFt{X,f}{t}$ for several values of $p$. The balls are colored differently according to their radius.
\end{minipage}
\hfill
\begin{minipage}[t]{0.30\textwidth}
\vspace{-10pt}
\begin{center}
    \includegraphics[width=.65\textwidth]{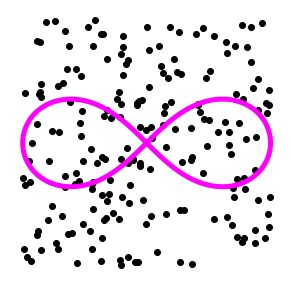}
  \end{center}
\end{minipage}

\noindent

The following proposition states the regularity of the persistence module $\V[X,f]$.% induced by $\CechF{X,f}$.
\begin{proposition}
\label{proposition_regularity_pers}
If $X \subseteq E$ is finite and $f$ is any function, then $\CechFpers{X,f}$ is a pointwise finite-dimensional persistence module. 

%More generally, if $X$ is any bounded subset of $E$ and $f$ is continuous, then $\CechFpers{X,f}$ is q-tame.
More generally, if $X$ is a bounded subset of $E$ and $f$ is any function, then $\CechFpers{X,f}$ is q-tame.
\end{proposition}

\begin{proof}
First, suppose that $X$ is finite. Then $\CechFnerve{X,f}$ is a filtration of a finite simplicial complex, and thus $\CechFnervepers{X,f}$ is pointwise finite-dimensional. It is also the case for $\CechFpers{X,f}$ since it is isomorphic to $\CechFnervepers{X,f}$. 

Secondly, suppose that $X$ is bounded. Consider the `filtration value' function:
\[ \begin{array}{lllll}
  t_X &: &E &\longrightarrow &\R^+ \\
   &&y  &\longmapsto &\inf \big\{ t \in \R^+, \exists x \in X, y \in \overline B_f(x,t) \big\}
\end{array} \]
For every $y\in E$, $x \in X$ and $t\geq 0$ the assertion $y \in \overline B_f(x,t)$ is equivalent to $\big( \|x-y\|^p+f(x)^p \big)^\frac{1}{p} \leq t$.
Therefore the function $t_X$ can be written as follows: 
\[t_X(y) = \inf \{\big( \|x-y\|^p+f(x)^p \big)^\frac{1}{p}, x \in X\}.\]
It is $1$-Lipschitz as it is the infimum of the set of the $1$-Lipschitz functions $y \mapsto \big(\|x-y\|^p+f(x)^p\big)^\frac{1}{p}$. It is also proper as $X$ is bounded. 

Let $\widetilde V$ be the filtration of $E$ defined for all $t \geq 0$ by $\widetilde V^t = t_X^{-1}(]-\infty, t])$. Let $\widetilde \V$ be its persistent homology module.
The last two properties of $t_X$ (continuous and proper) imply that $\widetilde \V$ is $q$-tame (\cite{Chazal_Persistencemodules}, Corollary 3.34).

Notice that, since $X$ may not be compact, $\CechFt{X,f}{t}$ may not be equal to $\widetilde V^t$. However, it follows from the definition of $t_X$ that $\CechF{X,f}$ and $\widetilde V$ are $\epsilon$-interleaved for every $\epsilon>0$. Therefore, $\CechFpers{X,f}$ also is $q$-tame.
\end{proof}

\subsection{Stability} 
We still consider a subset $X \subseteq E$ and a function $f\colon X \to \R^+$. 
Using the fact that two $\epsilon$-interleaved filtrations induce $\epsilon$-interleaved persistence modules, the stability results for the filtration $\CechF{X,f}$ of this subsection immediately translate as stability results for the persistence module $\CechFpers{X,f}$.

The following proposition relates the stability of the filtration $\CechF{X,f}$ with respect to $f$.

\begin{proposition}
Let $g\colon X \to \R^+$ be a function such that $\sup_{x\in X}|f(x) - g(x)| \leq \epsilon$. Then the filtrations $\CechF{X,f}$ and $\CechF{X,g}$ are $\epsilon$-interleaved.
\label{proposition_stability_1}
\end{proposition}

\begin{proof}
By symmetry, it suffices to show that, for every $t \geq 0$, $\CechFt{X,f}{t} \subseteq \CechFt{X,g}{t+\epsilon}$.  
\newline
Let $t\geq 0$. Choose $y \in \CechFt{X,f}{t}$, and $x\in X$ such that $y \in \overline B_f(x,t)$, i.e. $\big( \|x-y\|^p+f(x)^p \big)^\frac{1}{p} \leq t$. 
Let us prove that $y \in \overline B_g(x,t+\epsilon)$, i.e. $\big( \|x-y\|^p+g(x)^p \big)^\frac{1}{p} \leq t + \epsilon$. 
\newline
From $g(x) \leq f(x) + \epsilon$, we obtain $\big( \|x-y\|^p+g(x)^p \big)^\frac{1}{p} \leq \big( \|x-y\|^p+(f(x)+\epsilon)^p \big)^\frac{1}{p}$. 
Now, consider the function $\eta \mapsto \big( \|x-y\|^p+(f(x)+\eta)^p \big)^\frac{1}{p}$. Its derivative is $\eta \mapsto \Big(\frac{f(x)+\eta}{\big( \|x-y\|^p+(f(x)+\eta)^p \big)^\frac{1}{p}}\Big)^{p-1}$. It is consequently $1$-Lipschitz on $\R^+$. The Lipschitz property implies that
\[\big( \|x-y\|^p+\big(f(x)+\epsilon)^p\big)^\frac{1}{p} \leq \big( \|x-y\|^p+f(x)^p \big)^\frac{1}{p} + \epsilon.\]
Hence $\big( \|x-y\|^p+g(x)^p \big)^\frac{1}{p} \leq \big( \|x-y\|^p+(f(x)+\epsilon)^p \big)^\frac{1}{p} \leq \big( \|x-y\|^p+f(x)^p \big)^\frac{1}{p} + \epsilon \leq t +\epsilon$.
\end{proof}

The following proposition states the stability of $\CechF{X,f}$ with respect to $X$. It generalizes \cite[Proposition 4.3]{Buchet_Efficient} (case $p=2$).
% The $p=2$ case already appears in \cite{Buchet_Efficient}, Proposition 4.3.
%The following proposition is a generalization of \cite{Buchet_Thesis}, theorem 5.6.
\begin{proposition}
Let $Y \subseteq E$ and suppose that $f\colon X \cup Y \rightarrow \R^+$ is $c$-Lipschitz, $c \geq 0$. Suppose that $X$ and $Y$ are compact and that the Hausdorff distance $d_H(X,Y) \leq \epsilon$. Then the filtrations $\CechF{X,f}$ and $\CechF{Y,f}$ are $k$-interleaved with $k =\epsilon (1+c^p)^\frac{1}{p}$.
\label{proposition_stability_2}
\end{proposition}

\begin{proof}
It suffices to show that for every $t\geq 0$, $\CechFt{X,f}{t} \subseteq \CechFt{Y,f}{t + k}$. 
\newline \noindent
Let $t\geq 0$. Choose $z \in \CechFt{X,f}{t}$, and $x \in X$ such that $z \in \overline B_f(x, t)$, i.e. $\|x-z\|\leq r_x(t)$. From the hypothesis $d_H(X,Y) \leq \epsilon$, there exists $y \in Y$ such that $\|y-x\|\leq \epsilon$. Let us prove that $z \in \overline B_f(y, t+k)$, i.e. $\|z-y\|\leq r_y(t+k)$.

By triangle inequality, $\|z-y\| \leq \|z-x\| + \|x-y\| \leq r_x(t) + \epsilon$. It is enough to show that $r_x(t) + \epsilon \leq r_y(t+k)$, i.e. 
%For $f$ is $c$-Lipschitz, we have $f(y) \leq f(x) + c\|x-y\| \leq t + k$, thus the last inequality rephrases as
\[\underbrace{\big( (t+k)^p-f(y)^p \big)^\frac{1}{p}}_{r_y(t+k)}-\underbrace{\big( t^p-f(x)^p \big)^\frac{1}{p}}_{r_x(t)} \geq \epsilon .\]  
The left-hand side of this expression is decreasing in $f(y)$. Moreover, since $f$ is $c$-Lipschitz, $f(y)$ is at most $f(x)+c \epsilon$. Therefore:
\[ \begin{array}{ccccc}
& &((t+k)^p-f(y)^p)^\frac{1}{p} &- &(t^p-f(x)^p)^\frac{1}{p} \\ 
&\geq &((t+ k)^p-(f(x)+c\epsilon)^p)^\frac{1}{p} &- &(t^p-f(x)^p)^\frac{1}{p}.
\end{array} \]
It is now enough to prove that this last expression is not less than $\epsilon$, which is the content of Lemma \ref{lemma_inequality_stability}.
\end{proof}

Notice that the bounds in Proposition \ref{proposition_stability_1} and \ref{proposition_stability_2} are tight. 
In the first case, consider for example $E = \R$, the set $X = \{0\}$, and the functions $f = 0$ and $g = \epsilon$. For every $t < \epsilon$, we have $\CechFt{Y,f}{t} = \emptyset$, while $\CechFt{X,f}{t} \neq \emptyset$. 
Regarding the second proposition, consider $E = \R$, $f\colon x \mapsto cx$, $X = \{0\}$ and $Y = \{\epsilon\}$. We have, for every $t \geq 0$, $\CechFt{X,f}{t} = \overline B(0, t)$ and $\CechFt{Y,f}{t} = \overline B(\epsilon, (t^p - (c\epsilon)^p)^\frac{1}{p})$. For every $t < \epsilon (1+c^p)^\frac{1}{p}$, we have $(t^p - (c\epsilon)^p)^\frac{1}{p} < \epsilon$, hence $0 \notin \CechFt{Y,f}{t}$. In comparison, $\forall t \geq 0$, $0 \in \CechFt{X,f}{t}$.

\bigbreak
When considering data with outliers, the observed set $X$ may be very distant from the underlying signal $Y$ in Hausdorff distance. Therefore, the tight bound in Proposition \ref{proposition_stability_2} may be unsatisfactory. Moreover, a usual choice of $f$ would be $d_X$, the distance function to $X$. But the bound in Proposition \ref{proposition_stability_1} then becomes $\|d_X-d_Y\|_\infty = d_H(X,Y)$. 
We address this issue in Section \ref{section_DTM} by considering an outliers-robust function $f$, the so-called distance-to-measure function (DTM).

\subsection{Weighted Vietoris-Rips filtrations}
Rather than computing the persistence of the \v{C}ech filtration of a point cloud $X \subseteq E$, one sometimes consider the corresponding Vietoris-Rips filtration, which is usually easier to compute. 

If $G$ is a graph with vertex set $X$, its corresponding clique complex is the simplicial complex over $X$ consisting of the sets of vertices of cliques of $G$. If $S$ is a simplicial complex, its corresponding flag complex is the clique complex of its 1-skeleton. 
%We denote it $\Rips{S}$.

%We consider the case where $X$ is a finite point cloud. 
We remind the reader that $\CechFnervet{X,f}{t}$ denotes the nerve of $\CechFcovert{X,f}{t}$, where $\CechFcovert{X,f}{t}$ is the cover $\{\overline{B}_f(x, t)\}_{x \in X}$ of $\CechFt{X,f}{t}$. 

\begin{definition}
We denote by $\CechFripst{X,f}{t}$ the flag complex of $\CechFnervet{X,f}{t}$, and by $\CechFrips{X,f}$ the corresponding filtered simplicial complex. It is called the \it{weighted Rips complex with parameters $(X,f,p)$}. 
\label{definition_Rips}
\end{definition}

The following proposition states that the filtered simplicial complexes $\CechFnerve{X,f}$ and $\CechFrips{X,f}$ are 2-interleaved multiplicatively, generalizing the classical case of the \v{C}ech and Vietoris-Rips filtrations (case $f=0$).

\begin{proposition}
For every $t \geq 0$,
\[\CechFnervet{X,f}{t} \subseteq \CechFripst{X,f}{t} \subseteq \CechFnervet{X,f}{2t}\]
\end{proposition}

\begin{proof}
Let $t \geq 0$. The first inclusion follows from that $\text{Rips}(\CechFcovert{X,f}{t}))$ is the clique complex of $\CechFnervet{X,f}{t}$. To prove the second one, choose a simplex $\omega \in \text{Rips}(\CechFcovert{X,f}{t}))$. It means that for every $x, y \in \omega$, $\overline B_f(x,t) \cap \overline B_f(y, t) \neq \emptyset$, i.e. $\overline B(x,r_x(t)) \cap \overline B(y, r_y(t)) \neq \emptyset$. We have to prove that $\omega \in \CechFnervet{X,f}{2t}$, i.e. $\bigcap_{x \in \omega} \overline B(x, r_x(2t)) \neq \emptyset$. 

For every $x \in \omega$, one has $r_x(2t) \geq 2r_x(t)$. Indeed,
\begin{align*}
r_x(2t) &= \big( (2t)^p-f(x)^p \big)^\frac{1}{p} \\
&= 2 \big( t^p-(\frac{f(x)}{2})^p \big)^\frac{1}{p} \\
&\geq 2 \big( t^p-f(x)^p \big)^\frac{1}{p} = 2 r_x(t)
\end{align*}
Using the fact that doubling the radius of pairwise intersecting balls is enough to make them intersect globally, we obtain that $\omega \in \CechFnervet{X,f}{2t}$.
\end{proof}

%\begin{lemma}
%\label{lemma_intersecting_balls}
%Let $x_1, ..., x_n \in E$ and $r_1, ..., r_n > 0$. Suppose that the Euclidean balls $\overline B(x_i, r_i)$, $1\leq i \leq n$, intersect pairwise. Then
%\[\bigcap_{i=1}^n \overline B(x_i, 2 r_i) \neq \emptyset.\]
%\end{lemma}

%\begin{proof}
%Without loss of generality, assume that $r_1$ is the smallest radius. Let $i \in [2, n]$. From $\overline B(x_1, r_1) \cap \overline B(x_i, r_i) \neq \emptyset$ it follows that $x_1 \in \overline B(x_i, 2 r_i)$. So $x_1 \in \bigcap_{i=1}^n \overline B(x_i, 2 r_i)$. 
%\end{proof}

Using Theorem 3.1 of \cite{bell2017weighted}, the multiplicative interleaving $\CechFripst{X,f}{t} \subseteq \CechFnervet{X,f}{2t}$ can be improved to $\CechFripst{X,f}{t} \subseteq \CechFnervet{X,f}{ct}$, where $c = \sqrt{\frac{2d}{d+1}}$ and $d$ is the dimension of the ambient space $E = \R^d$. 

\bigbreak
Note that weighted Rips filtration shares the same stability properties as the weighted \v{C}ech filtration. Indeed, the proofs of Proposition \ref{proposition_stability_1} and \ref{proposition_stability_2} immediately extend to this case.
\bigbreak

In order to compute the flag complex $\CechFripst{X,f}{t}$, it is enough to know the filtration values of its 0- and 1-simplices. The following proposition describes these values.

\begin{proposition}
Let $p<+\infty$. The filtration value of a vertex $x \in X$ is given by $t_X(\{x\}) = f(x)$. 
\newline \noindent
The filtration value of an edge $\{x,y\} \subseteq E$ is given by 
\[ t_X(\{x,y\}) = \left\{ \begin{array}{cl}
\max \{f(x), f(y)\} & \text{ if } \|x-y\| \leq |f(x)^p - f(y)^p|^\frac{1}{p},  \\
t & \text{ otherwise,}
\end{array} \right. \]
where $t$ is the only positive root of 
\begin{equation}
\label{equation_edge_value}
\|x-y\| = (t^p - f(x)^p)^\frac{1}{p} + (t^p - f(y)^p)^\frac{1}{p}
\end{equation}
\label{prop_rips_values}
\end{proposition}

When $\|x-y\| \geq |f(x)^p - f(y)^p|^\frac{1}{p}$, the positive root of Equation \eqref{equation_edge_value} does not always admit a closed form. We give some particular cases for which it can be computed.
\begin{itemize}
\item For $p=1$, the root is $t_X(\{x,y\}) = \frac{f(x)+f(y)+\|x-y\|}{2}$,
\item for $p=2$, it is $t_X(\{x,y\}) = \frac{\sqrt{\big((f(x)+f(y))^2+\|x-y\|^2\big)\big((f(x)-f(y))^2+\|x-y\|^2\big)}}{2\|x-y\|}$,
\item for $p=\infty$, the condition reads $\|x-y\| \geq \max\{f(x), f(y)\}$, and the root is $t_X(\{x,y\}) = \frac{\|x-y\|}{2}$. In either case, $t_X(\{x,y\}) = \max\{f(x), f(y), \frac{\|x-y\|}{2} \}$.
\end{itemize}

\begin{proof}
The filtration value of a vertex $x \in X$ is, by definition of the nerve, $t_X(\{x\}) = \inf\{s \in T, \overline{B}_f(x, s) \neq \emptyset\}$. It is equal to $f(x)$.

Also by definition, the filtration value of an edge $\{x,y\}$, with $x,y \in X$ and $x \neq y$, is given by 
\[t_X(\{x,y\}) = \inf\{s \in \R, \overline{B}_f(x, s) \cap \overline{B}_f(y, s) \neq \emptyset\} \] 
Two cases may occur: the balls $\overline{B}_f(x, t(\{x,y\}))$ and $\overline{B}_f(x, t(\{x,y\}))$ have both positive radius, or one of these is a singleton. In the last case, $t(\{x,y\}) = \max \{f(x), f(y)\}$. In the first case, we have $\|x-y\| = r_x(t) + r_y(t)$, i.e. $\|x-y\| = (t^p - f(x)^p)^\frac{1}{p} + (t^p - f(y)^p)^\frac{1}{p}$. Notice that Equation \eqref{equation_edge_value} admits only one solution since the function $t \mapsto (t^p - f(x)^p)^\frac{1}{p} + (t^p - f(y)^p)^\frac{1}{p}$ is strictly increasing on $[\max\{f(x), f(y)\}, +\infty)$.
\end{proof}

We close this subsection by discussing the influence of $p$ on the weighted \v{C}ech and Rips filtrations. 
Let $D_0(\CechFnerve{X,f,p})$ be the persistence diagram of the 0th-homology of $\CechFnerve{X,f,p}$. We say that a point $(b,d)$ of $D_0(\CechFcover{X,f,p})$ is non-trivial if $b \neq d$. 
Let $D_0(\CechFrips{X,f,p})$ be the persistence diagram of the 0th-homology of $\CechFrips{X,f,p}$. Note that $D_0(\CechFnerve{X,f,p}) = D_0(\CechFrips{X,f,p})$ since the corresponding filtrations share the same 1-skeleton.

\begin{proposition}
The number of non-trivial points in $D_0(\CechFrips{X,f,p})$ is non-increasing with respect to $p \in [1, +\infty)$. The same holds for $D_0(\CechFnerve{X,f,p})$.
\label{prop_H0_nonincreasing}
\end{proposition}

\begin{proof}
%Since the persistence diagram of the 0th-homology of a filtered simplicial complex only depends on the filtration of its 1-skeleton, it is enough to show the lemma for $D_0(\CechFrips{X,f,p})$. As before, denote by $t_X$ the `filtration value' function of $\CechFrips{X,f,p}$.
The number of points in $D_0(\CechFrips{X,f,p})$ is equal to the cardinal of $X$. Any $p\geq 1$ being fixed, we can pair every $x \in X$ with some edge $\{y,z\} \in \CechFrips{X,f,p}$ such that the points of $D_0(\CechFrips{X,f,p})$ are of the form $\big(t_X(\{x\}), t_X(\{y,z\}) \big)$.

Notice that the filtration values of the points in $X$ do not depend on $p$: for all $p\geq 1$ and $x \in X$, $t_X(\{x\}) = f(x)$.
Moreover, the filtration values of the edges in $\CechFrips{X,f,p}$ are non-increasing with respect to $p$. Indeed, for all $\{y,z\} \in \CechFrips{X,f,p}$ with $y \neq z$, according to Proposition \ref{prop_rips_values}, the filtration value $t_X(\{y,z\})$ is either $\max \{f(x), f(y)\}$ if $\|x-y\| \leq |f(x)^p - f(y)^p|^\frac{1}{p}$, or is the only positive root of Equation \eqref{equation_edge_value} otherwise. 
Note that the positive root of Equation \eqref{equation_edge_value} is greater than $\max \{f(x), f(y)\}$ and decreasing in $p$. Besides, the term $|f(x)^p - f(y)^p|^\frac{1}{p}$ is non-decreasing in $p$.

These two facts ensure that for every $x \in X$, the point of $D_0(\CechFrips{X,f,p})$ created by $x$ has an ordinate which is non-increasing with respect to $p$. In particular, the number of non-trivial points in $D_0(\CechFrips{X,f,p})$ is non-increasing with respect to $p$.
\end{proof}

Figure \ref{Fig_diags_p>1} in Subsection \ref{subsection_p>1} illustrates the previous proposition in the case of the DTM-filtrations. Greater values of $p$ lead to sparser 0th-homology diagrams.

Now, consider $k>0$, and let $D_k(\CechFnerve{X,f,p})$ be the persistence diagram of the $k$th-homology of $\CechFnerve{X,f,p}$. In this case, one can easily build examples showing that the number of non-trivial points of $D_k(\CechFnerve{X,f,p})$ does not have to be non-increasing with respect to $p$. The same holds for $D_k(\CechFrips{X,f,p})$.

\section{DTM-filtrations}
\label{section_DTM}
The results of previous section suggest that in order to construct a weighted \v{C}ech filtration $\CechF{X,f}$ that is robust to outliers, it is necessary to choose a function $f$ that depends on $X$ and that is itself robust to outliers. The so-called distance-to-measure function (DTM) satisfies such properties, motivating the introduction of the DTM-filtrations in this section.
%The results of previous section suggests that in order to construct a weighted \v{C}ech filtration $\CechF{X,f}$ that is robust to outliers, it may be necessary to choose a function $f$ that depends on $X$ and that is itself robust to outliers. The so-called distance-to-measure function (DTM) satisfies such properties, motivating the introduction of the DTM-filtrations in this section. 

%As we have seen in the previous section, the stability of the weighted \v{C}ech filtration $\CechF{X,f}$ depends on the stability of $f$. Hence $f$ has to be robust to outliers. We choose $f$ to be the DTM of the empirical measure of the set $X$.

%Applying the results of the previous section, we immediately obtain a stability result (Proposition \ref{proposition_DTM-filtration_stability}), however including a term in Hausdorff distance. We then prove another kind of stability (Proposition \ref{prop_stability_p=1}), which does not involve Hausdorff distance.

%The choice of $p=1$ leads to the most robust filtration (Theorem \ref{prop_stability_p=1}). In this case, the DTM-filtration is close to the sublevel sets filtration of the DTM (Proposition \ref{prop_DTM_sublevel}). 

%When $p>1$, the stability cannot be expressed as an interleaving between set filtrations, but at the homological level (Theorem \ref{thm_p>1}).

\subsection{The distance to measure (DTM)}
%In this subsection we define the DTM and state some of its properties.

Let $\mu$ be a probability measure over $E = \R^d$, and $m \in [0,1)$ a parameter.
For every $x \in \R^d$, let $\delta_{\mu, m}$ be the function defined on $E$  by $\delta_{\mu, m}(x) = \inf \{r\geq0, \mu(\overline{B}(x, r))>m\}$. 

\begin{definition}
Let $m \in [0,1[$. The DTM $\mu$ of parameter $m$ is the function: 
\[\begin{array}{lrcl}
d_{\mu, m}\colon & E & \longrightarrow & \R \\
    & x & \longmapsto & \sqrt{ \frac{1}{m}\int_0^{m} \delta_{\mu,t}^2(x)dt}
\end{array} \]
When $m$ is fixed---which is the case in the following subsections---and when there is no risk of confusion, we write $d_{\mu}$ instead of $d_{\mu, m}$.
\end{definition}

Notice that when $m=0$, $d_{\mu,m}$ is the distance function to $\text{supp}(\mu)$, the support of $\mu$. 
%When $m >0$, it can be seen as a smoothing of it. 
%We state two results about the DTM.
\begin{proposition}[\cite{Chazal_Geometricinference}, Corollary 3.7]
For every probability measure $\mu$ and $m\in [0,1)$, $d_{\mu,m}$ is 1-Lipschitz.
\label{proposition_DTM_Lip}
\end{proposition}

A fundamental property of the DTM is its stability with respect to the probability measure $\mu$ in the Wasserstein metric. 
Given two probability measures $\mu$ and $\nu$ over $E$, a transport plan between $\mu$ and $\nu$ is a probability measure $\pi$ over $E \times E$ whose marginals are $\mu$ and $\nu$. The \textit{Wasserstein distance with quadratic cost between $\mu$ and $\nu$} is defined as $W_2(\mu,\nu) = \Big(\inf_\pi \int_{E\times E} \|x-y\|^2d\pi(x,y) \Big)^\frac{1}{2}$, where the infimum is taken over all the transport plans $\pi$.
When $\mu = \mu_X$ and $\nu = \mu_Y$ are the empirical measures of the finite point clouds $X$ and $Y$, i.e the normalized sums of the Dirac measures on the points of $X$ and $Y$ respectively, we write $W_2(X,Y)$ instead of $W_2(\mu_X, \mu_Y)$.

\begin{proposition}[\cite{Chazal_Geometricinference}, Theorem 3.5]
Let $\mu, \nu$ be two probability measures, and $m\in (0,1)$. Then
\begin{equation*}
\|d_{\mu,m} - d_{\nu,m}\|_\infty \leq m^{-\frac{1}{2}}W_2(\mu, \nu).
\end{equation*}

\label{proposition_DTM_norm}
\end{proposition}

Notice that for every $x \in E$, $d_{\mu}(x)$ is not lower than the distance from $x$ to $\text{supp}(\mu)$, the support of $\mu$. This remark, along with the propositions \ref{proposition_DTM_Lip} and \ref{proposition_DTM_norm}, are the only properties of the DTM that will be used to prove the results in the rest of the paper.
\bigbreak

In practice, the DTM can be computed. If $X$ is a finite subset of $E$ of cardinal $n$, we denote by $\mu_X$ its empirical measure. %$\mu_X = \frac{1}{n} \sum\limits_{i = 1}^n \delta_{x}$
Assume that $m = \frac{k_0}{n}$, with $k_0$ an integer.
In this case, $d_{\mu_X, m}$ reformulates as follows: for every $x \in E$,
\begin{align*}
d_{\mu_X, m}^2(x) = \frac{1}{k_0} \sum\limits_{k = 1}^{k_0} \|x-p_k(x)\|^2,
\end{align*}
where $p_1(x), ..., p_{k_0}(x)$ are a choice of $k_0$-nearest neighbors of $x$ in $X$.

\subsection{DTM-filtrations}
In the following, the two parameters $p \in [1, +\infty]$ and $m \in (0,1)$ are fixed.

\begin{definition}
Let $X\subseteq E$ be a finite point cloud, $\mu_X$ the empirical measure of $X$, and $d_{\mu_X}$ the corresponding DTM of parameter $m$.
The weighted \v{C}ech filtration $\CechF{X,d_{\mu_X}}$, as defined in Definition \ref{def_weightedCech}, is called the \emph{DTM-filtration associated with the parameters $(X, m, p)$}. 
It is denoted by $\DTMF{X}$. The corresponding persistence module is denoted by $\DTMFpers{X}$.
\end{definition}

%In the sequel, we use the simplicial counterpart of $\DTMF{X}$. 
Let $\DTMFcovert{X}{t} = \CechFcovert{X,d_{\mu_X}}{t}$ denote the cover of $\DTMFt{X}{t}$ as defined in section \ref{section_Vp_modules}, and let $\DTMFnervet{X}{t}$ be its nerve. 
The family $\DTMFnerve{X}) = (\DTMFnervet{X}{t})_{t\geq 0}$ is a filtered simplicial complex, and its persistent (simplicial) homology module is denoted by $\DTMFnervepers{X}$. By the persistent nerve lemma, the persistence modules $\DTMFpers{X}$ and $\DTMFnervepers{X}$ are isomorphic.

As in Definition \ref{definition_Rips}, $\DTMFripst{X}{t}$ denotes the flag complex of $\DTMFnervet{X}{t}$, and $\DTMFrips{X}$ the corresponding filtered simplicial complex.

\begin{figure}[H]
   \centering
\begin{tabular}{ccc}
\includegraphics[width=4cm]{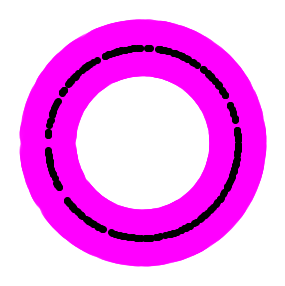}&
\includegraphics[width=3.8cm]{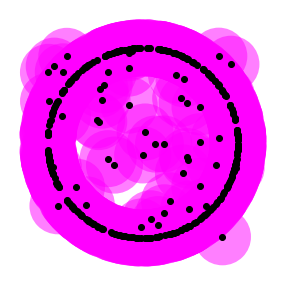}&
\includegraphics[width=4cm]{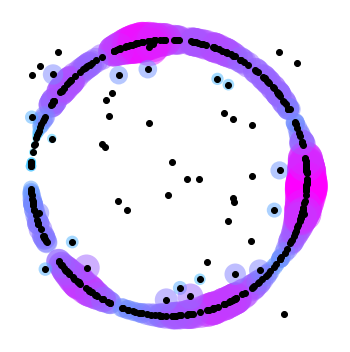}\\
$\UsualCechFt{\Gamma}{t}$ & $\UsualCechFt{X}{t}$ & $\DTMFt{X}{t}$\\
\end{tabular}
\label{Fig_CechvsDTM}
\caption{The sets $\UsualCechFt{\Gamma}{t}$, $\UsualCechFt{X}{t}$ and $\DTMFt{X}{t}$ for $p=1$, $m=0{,}1$ and $t=0{,}3$.}
\end{figure}

\bigbreak
\noindent
\begin{minipage}[t]{0.74\textwidth}
\begin{example} 
Consider the point cloud $X$ drawn on the right. It is the union of $\widetilde{X}$ and $\Gamma$, where $\widetilde X$ is a 50-sample of the uniform distribution on $[-1,1]^2 \subseteq \R^2$, and $\Gamma$ is a 300-sample of the uniform distribution on the unit circle. 
We consider the weighted \v{C}ech filtrations $\UsualCechF{\Gamma}$ and $\UsualCechF{X}$, and the DTM-filtration $\DTMF{X}$, for $p=1$ and $m=0{,}1$. They are represented in Figure \ref{Fig_CechvsDTM} for the value $t=0{,}3$.
\label{example_CechvsDTM}
\end{example}
\end{minipage}
\hfill
\begin{minipage}[t]{0.25\textwidth}
\vspace{-10pt}
\begin{center}
    \includegraphics[width=.7\textwidth]{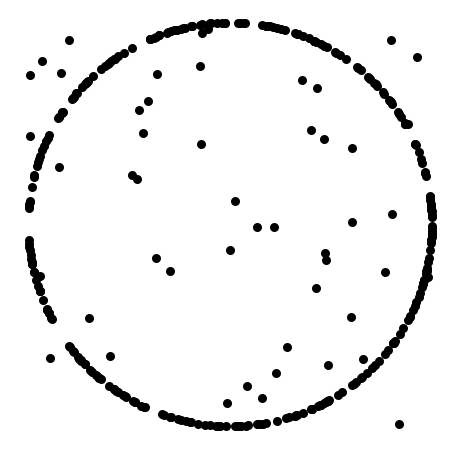}
  \end{center}
\end{minipage}

Because of the outliers $\widetilde X$, the value of $t$ from which the sets $\UsualCechFt{X}{t}$ are contractible is small. On the other hand, we observe that the set $\DTMFt{X}{t}$ does not suffer too much from the presence of outliers. 

\begin{figure}[H]
   \centering
\begin{tabular}{ccc}
\includegraphics[width=4.5cm]{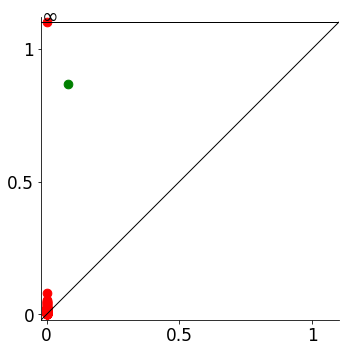}&
\includegraphics[width=4.5cm]{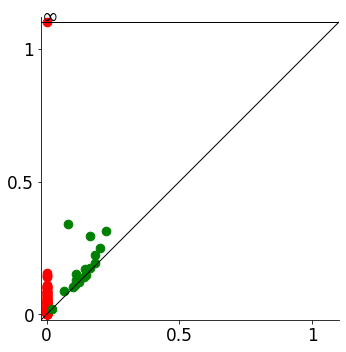}&
\includegraphics[width=4.5cm]{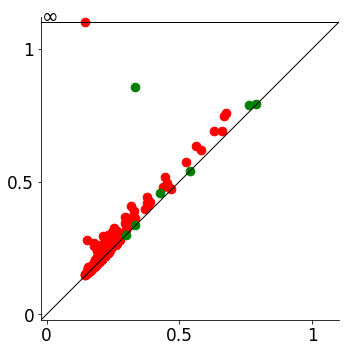}\\
$D(\CechFrips{\Gamma, 0})$ & $D(\CechFrips{X, 0})$ & $D(\DTMFrips{X})$ \\
\end{tabular}
\caption{Persistence diagrams of some simplicial filtrations. Points in red (resp. green) represent the persistent homology in dimension $0$ (resp. $1$).}
\label{Fig_diags_p=1}
\end{figure}

We plot in Figure \ref{Fig_diags_p=1} the persistence diagrams of the persistence modules associated to $\CechFrips{\Gamma, 0}$, $\CechFrips{X, 0}$ and $\DTMFrips{X}$ ($p=1$, $m=0{,}1$).

Observe that the diagrams $D(\CechFrips{\Gamma, 0})$ and $D(\DTMFrips{X})$ appear to be close to each other, while $D(\CechFrips{X, 0})$ does not. 
\bigbreak

Applying the results of Section \ref{section_Vp_modules}, we immediately obtain the following proposition.

\begin{proposition}
\label{proposition_DTM-filtration_stability}
Consider two measures $\mu, \nu$ on $E$ with compact supports $X$ and $Y$. Then
\[d_i(\CechF{X, d_{\mu}}, \CechF{Y, d_{\nu}}) \leq m^{-\frac{1}{2}}W_2(\mu, \nu) + 2^\frac{1}{p} d_H(X,Y). \]
In particular, if $X$ and $Y$ are finite subsets of $E$, using $\mu = \mu_X$ and $\nu = \nu_Y$, we obtain
\[d_i(\DTMF{X}, \DTMF{Y}) \leq m^{-\frac{1}{2}}W_2(X, Y) + 2^\frac{1}{p} d_H(X,Y). \]
\end{proposition}

\begin{proof}
We use the triangle inequality for the interleaving distance:
\[d_i(\CechF{X, d_{\mu}}, \CechF{Y, d_{\nu}}) \leq \underbrace{d_i(\CechF{X, d_{\mu}}, \CechF{Y, d_{\mu}})}_{(1)} + \underbrace{d_i(\CechF{Y, d_{\mu}}, \CechF{Y, d_{\nu}})}_{(2)}.\] 

\paragraphe{(1)} From Proposition \ref{proposition_stability_2}, we have $d_i(\CechF{X, d_{\mu}}, V[Y, d_{\mu}]) \leq (1+c^p)^\frac{1}{p} d_H(X,Y)$, where $c$ is the Lipschitz constant of $d_{\mu}$. According to Proposition \ref{proposition_DTM_Lip}, $c=1$. We obtain $d_i(\CechF{X, d_{\mu}}, V[Y, d_{\mu}]) \leq 2^\frac{1}{p} d_H(X,Y)$.

\paragraphe{(2)} From Proposition \ref{proposition_stability_1}, we have $d_i(\CechF{Y, d_{\mu}}, \CechF{Y, d_{\nu}}) \leq \|d_{\mu}-d_{\nu}\|_\infty$. According to Proposition \ref{proposition_DTM_norm}, $\|d_{\mu}-d_{\nu}\|_\infty \leq m^{-\frac{1}{2}} W_2(\mu, \nu)$. 

%We use the triangle inequality for the interleaving distance:
%\[d_i(\CechF{X, d_{\mu_X}}, \CechF{Y, d_{\mu_Y}}) \leq \underbrace{d_i(\CechF{X, d_{\mu_X}}, \CechF{Y, d_{\mu_X}})}_{(1)} + \underbrace{d_i(\CechF{Y, d_{\mu_X}}, \CechF{Y, d_{\mu_Y}})}_{(2)}.\] 
%\paragraphe{(1)} From Proposition \ref{proposition_stability_2}, we have $d_i(\CechF{X, d_{\mu_X}}, V[Y, d_{\mu_X}]) \leq (1+c^p)^\frac{1}{p} d_H(X,Y)$, where $c$ is the Lipschitz constant of $d_{\mu_X}$. According to Proposition \ref{proposition_DTM_Lip}, $c=1$. We obtain $d_i(\CechF{X, d_{\mu_X}}, V[Y, d_{\mu_X}]) \leq 2^\frac{1}{p} d_H(X,Y)$.
%\paragraphe{(2)} From Proposition \ref{proposition_stability_1}, we have $d_i(\CechF{Y, d_{\mu_X}}, \CechF{Y, d_{\mu_Y}}) \leq \|d_{\mu_X}-d_{\mu_Y}\|_\infty$. According to Proposition \ref{proposition_DTM_norm}, $\|d_{\mu_X}-d_{\mu_Y}\|_\infty \leq m^{-\frac{1}{2}} W_2(X, Y)$. 
\medbreak
The second point follows from the definition of the DTM-filtrations: $\DTMF{X} = \CechF{X, d_{\mu_X}}$ and $\DTMF{Y} = \CechF{Y, d_{\mu_Y}}$
\end{proof}

Note that this stability result is worse than the stability of the usual \v{C}ech filtrations, which only involves the Hausdorff distance. However, the term $W_2(X, Y)$ is inevitable, as shown in the following example. 

Let $E = \R$, and $\epsilon \in (0,1)$. 
Define $\mu = \epsilon \delta_0 + (1-\epsilon) \delta_1$, and $\nu = (1-\epsilon) \delta_0 + \epsilon \delta_1$. We have $X = \mathrm{supp}{(\mu)} = \mathrm{supp}{(\nu)} = Y$.
If $\epsilon \leq m \leq 1-\epsilon$, then $d_{\nu}(0) = 0$, while $d_{\mu}(0) = \sqrt{1-\frac{\epsilon}{m}}$.
We deduce that $d_i(\CechF{X, d_{\mu}}, \CechF{Y, d_{\nu}}) \geq d_{\mu}(0)-d_{\nu}(0) = \sqrt{1-\frac{\epsilon}{m}}$.

In comparison, the usual \v{C}ech filtrations $\CechF{X, 0}$ and $\CechF{Y, 0}$ are equal and does not depend on $\mu$ and $\nu$.
In this case, it would be more robust to consider these usual \v{C}ech filtrations.
Now, in the case where the Hausdorff distance $d_H(X,Y)$ is large, the usual \v{C}ech filtrations may be very distant. However, the DTM-filtrations may still be close, as we discuss in the next subsection.

%Pick two finite subsets $X$ and $Y$ of $E$ close to $Z$ in Hausdorff distance, and such that $\mu_X$ is close to $\epsilon \delta_0 + (1-\epsilon) \delta_1$ in Wasserstein distance, and $\mu_Y$ close to $(1-\epsilon) \delta_0 + \epsilon \delta_1$. 
%If $m\geq \epsilon$, then $d_{\mu_Y}(0)$ is close to $0$, while $d_{\mu_X}(0)$ is close to $\sqrt{1-\frac{\epsilon}{m}}$.

%If $p=1$, the interleaving distance $d_i(\DTMF{X}, \DTMF{Y})$ is then close to $\sqrt{1-\frac{\epsilon}{m}}$ (as in Proposition \ref{proposition_stability_1}).

%In comparison, the interleaving distance between the usual \v{C}ech filtrations is close to 0.
%In this case, it would be more robust to consider these usual \v{C}ech filtrations.
%However, in the case where the Hausdorff distance $d_H(X,Y)$ is large, the usual \v{C}ech filtrations may be very distant. On the other hand, the DTM-filtrations may still be close, as we discuss in the next subsection.

\subsection{Stability when \texorpdfstring{$p=1$}{p=1}}
\label{subsection_p=1}
We first consider the case $p=1$, for which the proofs are simpler and the results stronger.
\medbreak
We fix $m \in (0,1)$. If $\mu$ is a probability measure on $E$ with compact support $\text{supp}(\mu)$, we define
\[ c(\mu, m) = \sup_{\text{supp}(\mu)}(d_{\mu,m}). \]
If $\mu = \mu_{\Gamma}$ is the empirical measure of a finite set $\Gamma \subseteq E$, we denote it $c(\Gamma,m)$.

\begin{proposition}
Let $\mu$ be a probability measure on $E$ with compact support $\Gamma$. Let $d_{\mu}$ be the corresponding DTM. Consider a set $X \subseteq E$ such that $\Gamma \subseteq X$. The weighted \v{C}ech filtrations $\CechF{\Gamma, d_{\mu}}$ and $\CechF{X, d_{\mu}}$ are $c(\mu, m)$-interleaved.

Moreover, if $Y \subseteq E$ is another set such that $\Gamma \subseteq Y$, $\CechF{X, d_{\mu}}$ and $\CechF{Y, d_{\mu}}$ are $c(\mu, m)$-interleaved.

In particular, if $\Gamma$ is a finite set and $\mu = \mu_\Gamma$ its empirical measure, $\DTMF{\Gamma}$ and $\CechF{X, d_{\mu_\Gamma}}$ are $c(\Gamma, m)$-interleaved.
\label{prop_cratio}
\end{proposition}

\begin{proof}
Let $c=c(\mu, m)$. 
Since $\Gamma \subseteq X$, we have $\CechFt{\Gamma, d_\mu}{t} \subseteq \CechFt{X, d_\mu}{t}$ for every $t \geq 0$. 

Let us show that, for every $t \geq 0$, $\CechFt{X, d_\mu}{t} \subseteq \CechFt{\Gamma, d_\mu}{t+c}$.
Let $x \in X$, and choose $\gamma \in \Gamma$ a projection of $x$ on the compact set $\Gamma$, i.e. one of the closest points to $x$ in $\Gamma$. By definition of the DTM, we have that $d_\mu(x) \geq \|x-\gamma\|$.
Together with $d_\mu(\gamma) \leq c$, we obtain
\[t+c-d_\mu(\gamma) \geq t \geq t - d_\mu(x) + \|x-\gamma\|,\]
which means that $\overline B_{d_\mu}(x,t) \subseteq \overline B_{d_\mu}(\gamma, t+c)$. The inclusion $\CechFt{X, d_\mu}{t} \subseteq \CechFt{\Gamma, d_\mu}{t+c}$ follows. 

\smallbreak
If $Y$ is another set containing $\Gamma$, we obtain $\CechFt{X, d_\mu}{t} \subseteq \CechFt{\Gamma, d_\mu}{t+c} \subseteq \CechFt{Y, d_\mu}{t+c}$ for every $t \geq 0$.
\end{proof}

\begin{theorem}
Consider two measures $\mu, \nu$ on $E$ with supports $X$ and $Y$. Let $\mu', \nu'$ be two measures with compact supports $\Gamma$ and $\Omega$ such that $\Gamma \subseteq X$ and $\Omega \subseteq Y$.
We have
\[ d_i(\CechF{X,d_\mu},\CechF{Y,d_\nu}) \leq 
m^{-\frac{1}{2}}W_2(\mu,\mu') + m^{-\frac{1}{2}}W_2(\mu',\nu')  + m^{-\frac{1}{2}}W_2(\nu',\nu) + c(\mu',m) + c(\nu',m).\]
In particular, if $X$ and $Y$ are finite, we have
\[ d_i(\DTMF{X},\DTMF{Y}) \leq 
m^{-\frac{1}{2}}W_2(X,\Gamma) + m^{-\frac{1}{2}}W_2(\Gamma,\Omega) + m^{-\frac{1}{2}}W_2(\Omega,Y) + c(\Gamma,m) + c(\Omega,m).\]
Moreover, with $\Omega = Y$, we obtain
\[ d_i(\DTMF{X},\DTMF{\Omega}) \leq 
m^{-\frac{1}{2}}W_2(X,\Gamma) + m^{-\frac{1}{2}}W_2(\Gamma,\Omega) + c(\Gamma,m) + c(\Omega,m).\]

\label{prop_stability_p=1}
\end{theorem}

\begin{proof}%[Proof of theorem \ref{prop_stability_p=1}]
Let $d_X = d_{\mu}$, $d_Y = d_{\nu}$, $d_\Gamma = d_{\mu'}$ and $d_\Omega = d_{\nu'}$. 
We prove the first assertion by introducing the following filtrations between $\CechF{X, d_X}$ and $\CechF{Y, d_Y}$:
\[\CechF{X, d_X} \longleftrightarrow \CechF{X, d_\Gamma} \longleftrightarrow \CechF{\Gamma \cup \Omega, d_\Gamma} \longleftrightarrow \CechF{\Gamma \cup \Omega, d_\Omega} \longleftrightarrow \CechF{Y, d_\Omega} \longleftrightarrow \CechF{Y, d_Y}. \]
We have:
\begin{align*}
&\bullet d_i(\CechF{X, d_X} , \CechF{X, d_\Gamma}) \leq m^{-\frac{1}{2}}W_2(\mu,\mu')
&& \text{(Propositions \ref{proposition_DTM_norm} and \ref{proposition_stability_1})}, \\
&\bullet d_i(\CechF{X, d_\Gamma} , \CechF{\Gamma \cup \Omega, d_\Gamma})\leq c(\mu',m)
&& \text{(Proposition \ref{prop_cratio})}, \\
&\bullet d_i(\CechF{\Gamma \cup \Omega, d_\Gamma}, \CechF{\Gamma \cup \Omega, d_\Omega})\leq m^{-\frac{1}{2}}W_2(\mu',\nu')
&& \text{(Propositions \ref{proposition_DTM_norm} and \ref{proposition_stability_1})}, \\
&\bullet d_i(\CechF{\Gamma \cup \Omega, d_\Omega} , \CechF{Y, d_\Omega}) \leq c(\nu', m)
&& \text{(Proposition \ref{prop_cratio})}, \\
&\bullet d_i(\CechF{Y, d_\Omega} , \CechF{Y, d_Y}) \leq m^{-\frac{1}{2}}W_2(\nu', \nu)
&& \text{(Propositions \ref{proposition_DTM_norm} and \ref{proposition_stability_1})}.
\end{align*}

The inequality with $X$ and $Y$ finite follows from defining $\mu, \nu, \mu'$ and $\nu'$ to be the empirical measures on $X,Y,\Gamma$ and $\Omega$, and by recalling that the DTM filtrations $\DTMF{X}$ and $\DTMF{Y}$ are equal to the weighted \v{C}ech filtration $\CechF{X, d_\mu}$ and $\CechF{Y, d_\nu}$.
\end{proof}

The last inequality of Theorem \ref{prop_stability_p=1} can be seen as an approximation result. Indeed, suppose that $\Omega$ is some underlying set of interest, and $X$ is a sample of it with, possibly, noise or outliers. If one can find a subset $\Gamma$ of $X$ such that $X$ and $\Gamma$ are close to each other---in the Wasserstein metric---and such that $\Gamma$ and $\Omega$ are also close, then the filtrations $\DTMF{X}$ and $\DTMF{\Omega}$ are close. Their closeness depends on the constants $c(\Gamma,m)$ and $c(\Omega,m)$.
More generally, if $X$ is finite and $\mu'$ is a measure with compact support $\Omega \subset X$ not necessarily finite, note that the first inequality gives
\[ d_i(\DTMF{X},\CechF{\Omega, d_{\mu'}}) \leq 
m^{-\frac{1}{2}}W_2(X,\Gamma) + m^{-\frac{1}{2}}W_2(\mu_\Gamma,\mu') + c(\Gamma,m) + c(\mu',m).\]

For any probability measure $\mu$ of support $\Gamma \subseteq E$, the constant $c(\mu,m)$ might be seen as a bias term, expressing the behaviour of the DTM over $\Gamma$. It relates to the concentration of $\mu$ on its support. A usual case is the following: a measure $\mu$ with support $\Gamma$ is said to be $(a,b)$-standard, with $a,b \geq 0$, if for all $x \in \Gamma$ and $r \geq 0$, $\mu(\overline B(x,r)) \geq \min \{ ar^b, 1\}$. 
For example, the Hausdorff measure associated to a compact $b$-dimensional submanifold of $E$ is $(a,b)$-standard for some $a>0$.
In this case, a simple computation shows that there exists a constant $C$, depending only on $a$ and $b$, such that for all $x \in \Gamma$, $d_{\mu,m}(x)\leq C m^\frac{1}{b}$. Therefore, $c(\mu,m) \leq C m^\frac{1}{b}$. 

Regarding the second inequality of Theorem \ref{prop_stability_p=1}, suppose for the sake of simplicity that one can choose $\Gamma = \Omega$. The bound of Theorem \ref{prop_stability_p=1} then reads
\[ d_i(\DTMF{X},\DTMF{Y}) \leq 
m^{-\frac{1}{2}}W_2(X,\Gamma) + m^{-\frac{1}{2}}W_2(\Gamma,Y) + 2c(\Gamma,m).\]
Therefore, the DTM-filtrations $\DTMF{X}$ and $\DTMF{Y}$ are close to each other if $\mu_X$ and $\mu_Y$ are both close to a common measure $\mu_\Gamma$. This would be the case if $X$ and $Y$ are noisy samples of $\Gamma$.
This bound, expressed in terms of Wasserstein distance rather than Hausdorff distance, shows the robustness of the DTM-filtration to outliers.

\smallbreak
Notice that, in practice, for finite data sets $X, Y$ and for given $\Gamma$ and $\Omega$, the constants $c(\Gamma,m)$ and $c(\Omega,m)$ can be explicitly computed, as it amounts to evaluating the DTM on $\Gamma$ and $\Omega$. This remark holds for the bounds of Theorem \ref{prop_stability_p=1}.

\bigbreak \noindent
\begin{example}
%We illustrate Proposition \ref{prop_cratio} and Theorem \ref{prop_stability_p=1}. 
Consider the set $X = \widetilde X \cup \Gamma$ as defined in the example page \pageref{example_CechvsDTM}.
Figure \ref{Fig_ex_p=1} displays the sets $\DTMFt{X}{t}$, $\CechFt{X,d_{\mu_\Gamma}}{t}$ and $\DTMFt{\Gamma}{t}$ for the values $p=1$, $m=0.1$ and $t=0.4$ and the persistence diagrams of the corresponding weighted Rips filtrations, illustrating the stability properties of Proposition \ref{prop_cratio} and Theorem \ref{prop_stability_p=1}.

\begin{figure}[H]
   \centering
\begin{tabular}{ccc}
\includegraphics[width=3.5cm]{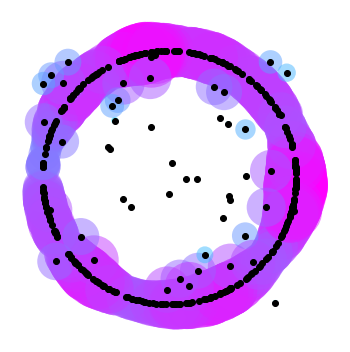}&
\includegraphics[width=3.5cm]{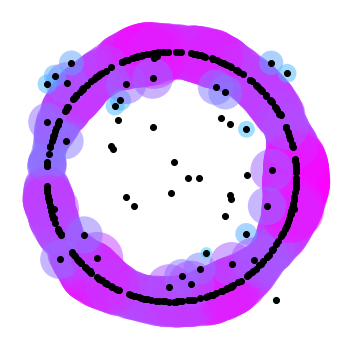}&
\includegraphics[width=3.5cm]{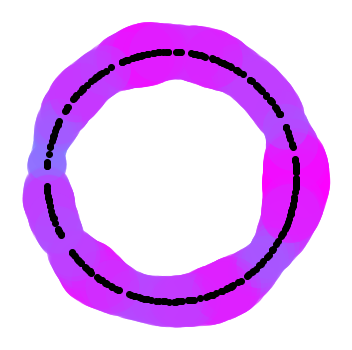}\\
$\DTMFt{X}{t}$ & $\CechFt{X,d_{\mu_\Gamma}}{t}$ & $\DTMFt{\Gamma}{t}$\\
\includegraphics[width=4.5cm]{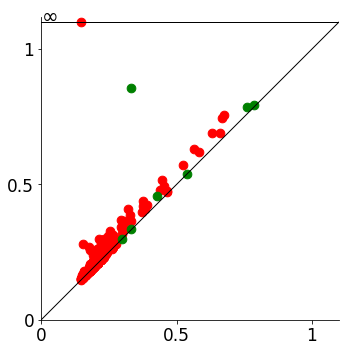}&
\includegraphics[width=4.5cm]{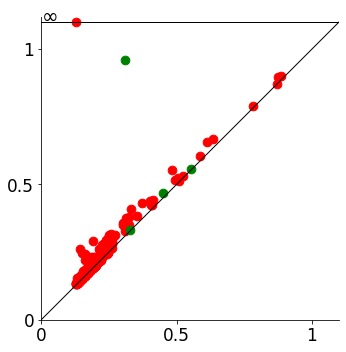}&
\includegraphics[width=4.5cm]{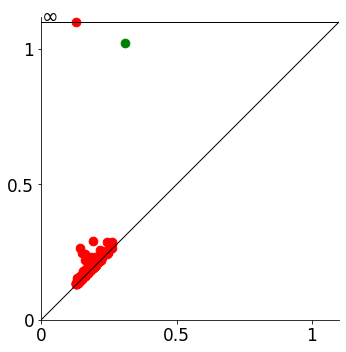}\\
$D(\DTMFrips{X})$ & $D(\CechFrips{X,d_{\mu_\Gamma}})$ & $D(\DTMFrips{\Gamma})$\\
\end{tabular}
\caption{Filtrations for $t=0.4$, and their corresponding persistence diagrams.}
\label{Fig_ex_p=1} 
\end{figure}
\end{example}
\bigbreak

The following proposition relates the DTM-filtration to the filtration of $E$ by the sublevels sets of the DTM.

\begin{proposition} \label{prop:DTMvsDTMfilt}
Let $\mu$ be a probability measure on $E$ with compact support $K$. Let $m \in [0,1)$ and denote by $V$ the sublevel sets filtration of $d_{\mu}$. Consider a finite set $X \subseteq E$. Then
\[d_i(V, \DTMF{X}) \leq m^{-\frac{1}{2}}W_2(\mu, \mu_X) + 2\epsilon + c(\mu, m),\]
with $\epsilon = d_H(K\cup X,X)$.
\label{prop_DTM_sublevel}
\end{proposition}

\begin{proof}
First, notice that $V = V[E, d_\mu]$. Indeed, for every $t \geq 0$, we have $V^t \subseteq \CechFt{E, d_\mu}{t}$ by definition of the weighted \v{C}ech filtration. To prove that $\CechFt{E, d_\mu}{t} \subseteq V^t$, let $x \in \CechFt{E, d_\mu}{t}$, and $y \in E$ such that $x \in \overline B_{d_\mu}(y,t)$. We have $\|x-y\| \leq t - f(y)$. For $d_\mu$ is 1-Lipschitz, we deduce $f(x) \leq f(y) + \|x-y\| \leq f(y) + t - f(y) \leq t$. Hence $x \in V^t$.
\smallbreak
Then we compute:
\begin{align*}
 d_i(V, \DTMF{X}) &= d_i(\CechF{E, d_\mu}, \CechF{X, d_{\mu_X}}) \\
& \leq d_i(\CechF{E, d_{\mu}}, \CechF{X \cup K, d_{\mu }}) + d_i(\CechF{X\cup K, d_{\mu}}, \CechF{X, d_{\mu}}) + d_i(\CechF{X, d_{\mu}}, \CechF{X, d_{\mu_X}}) \\
& \leq c(\mu,m) + 2\epsilon + m^{-\frac{1}{2}}W_2(\mu, \mu_X),
\end{align*}
using Proposition \ref{proposition_DTM_norm} for the first term, Proposition \ref{proposition_stability_2} for the second one, and Proposition \ref{proposition_stability_1} and Proposition \ref{prop_cratio} for the third one.
\end{proof}

As a consequence, one can use the DTM-filtration to approximate the persistent homology of the sublevel sets filtration of the DTM, which is expensive to compute in practice.

\medbreak
We close this subsection by noting that a natural strengthening of Theorem \ref{prop_stability_p=1} does not hold: let $m \in (0,1)$ and $E = \R^n$ with $n \geq 1$. There is no constant $C$ such that, for every probability measure $\mu, \nu$ on $E$ with supports $X$ and $Y$, we have:
\[ d_i(\CechF{X,d_{\mu,m}},\CechF{Y,d_{\nu,m}}) \leq C W_2(\mu,\nu).\]
The same goes for the weaker statement
\[ d_i(\CechFpers{X,d_{\mu,m}},\CechFpers{Y,d_{\nu,m}}) \leq C W_2(\mu,\nu).\]

We shall prove the statement for $E=\R$. 
%The cases $m>\frac{1}{2}$ can be proven with a similar construction.
Let $q \in (0,1)$ such that $q<m<1-q$, and $\epsilon \in [0,q)$. Let $x \in (-1,0)$ to be determined later. 
Define $\mu = q\delta_{-1} + (1-q)\delta_{1}$, and $\nu^{\epsilon} = (q-\epsilon)\delta_{-1} + (1-q)\delta_{1} + \epsilon \delta_x$, with $\delta$ denoting the Dirac mass.
Let $X = \{-1, 1\} \subset E$ and $Y = \{-1, x, 1\}$.

It is clear that $W_2(\mu,\nu^{\epsilon}) = (x+1) \epsilon < \epsilon$. 
Using the triangle inequality, we have:
\begin{align*}
d_i(\CechFpers{X,d_{\mu,m}},\CechFpers{Y,d_{\nu^{\epsilon},m}})
&\geq d_i(\CechFpers{X,d_{\mu,m}},\CechFpers{Y,d_{\mu,m}}) - d_i(\CechFpers{Y,d_{\nu^{\epsilon},m}},\CechFpers{Y,d_{\mu,m}}) \\
&\geq d_i(\CechFpers{X,d_{\mu,m}},\CechFpers{Y,d_{\mu,m}}) - m^{-\frac{1}{2}}\epsilon
\end{align*}
Thus it is enough to show that $d_i(\CechFpers{X,d_{\mu,m}},\CechFpers{Y,d_{\mu,m}})$ is positive.

%We shall show that, as $\epsilon$ goes to 0, $d_i(\CechFpers{X,d_{\mu,m}},\CechFpers{Y,d_{\nu^{\epsilon},m}})$ goes to a positive constant. 
%In fact, the persistence diagram of $\CechFpers{Y,d_{\nu^{\epsilon},m}}$ goes to the one of $\CechFpers{Y,d_{\mu,m}}$, and $d_i(\CechFpers{X,d_{\mu,m}},\CechFpers{Y,d_{\mu,m}})$ is positive.

Since $1-q>m$, we have $d_{\mu,m}(1) = 0$. 
%Moreover, $d_{\mu,m}(-1) = 2 \sqrt{\frac{m-q}{m}}$. 
Using Proposition \ref{prop_rips_values}, we deduce that the persistence barcode of the 0th homology of $\CechF{X,d_\mu}$ consists of the bars $[0, +\infty[$ and $[d_{\mu,m}(-1), \frac{1}{2}(d_{\mu,m}(-1)+d_{\mu,m}(1)+2)]$.% = [2 \sqrt{\frac{m-q}{m}},\sqrt{\frac{m-q}{m}}+1]$.

Similarly, the persistence barcode of the 0th homology of $\CechF{Y,d_{\mu}}$ consists of the bars $[0, +\infty[$, 
$[d_{\mu,m}(-1), \frac{1}{2}(d_{\mu,m}(-1)+d_{\mu,m}(x)+(1+x))]$
and $[d_{\mu,m}(x), \frac{1}{2}(d_{\mu,m}(x)+(1-x))]$.

Notice that, since $q>0$ and $x<0$, by definition of the DTM, we have $d_{\mu,m}(x) < 1-x$. Hence the last bar is not a singleton. 
Moreover, if $x$ is close enough to 0, we have $d_{\mu,m}(-1) < d_{\mu,m}(x)+1-x$. Indeed, with $x=0$, $d_{\mu,m}(x)+1-x = 2$, and we have $d_{\mu,m}(-1) = 2 \sqrt{\frac{m-q}{m}} < 2$. Hence the second bar is not a singleton as well.

As a consequence, if $x$ is close enough to 0, the interleaving distance between these two barcodes is positive.

\subsection{Stability when \texorpdfstring{$p>1$}{p>1}}
\label{subsection_p>1}
Now assume that $p > 1$, $m \in (0,1)$ being still fixed.
In this case, stability properties turn out to be more difficult to establish. For small values of $t$, Lemma \ref{lemma_DTMF_mult} gives a tight non-additive interleaving between the filtrations. However, for large values of $t$, the filtrations are poorly interleaved. To overcome this issue we consider stability at the homological level, i.e. between the persistence modules associated to the DTM filtrations. 

\begin{figure}[H]
   \centering
\begin{tabular}{ccc}
\includegraphics[width=3cm]{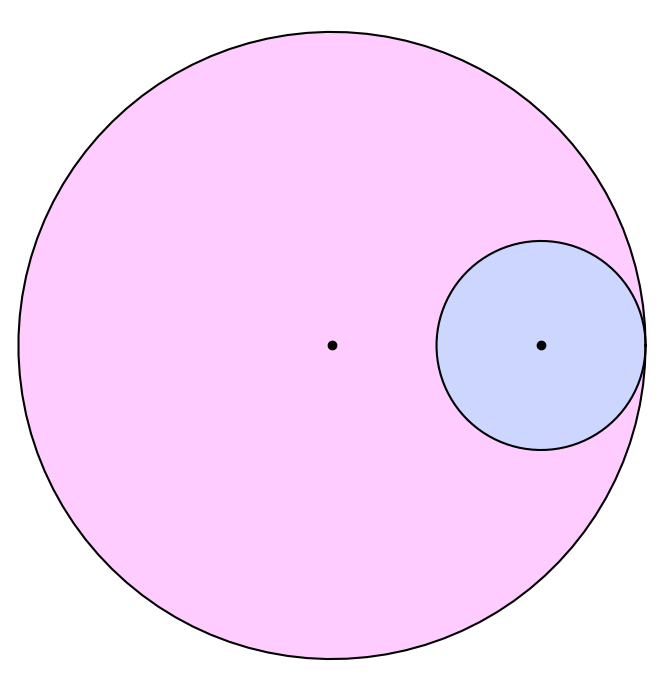}&
\includegraphics[width=3.5cm]{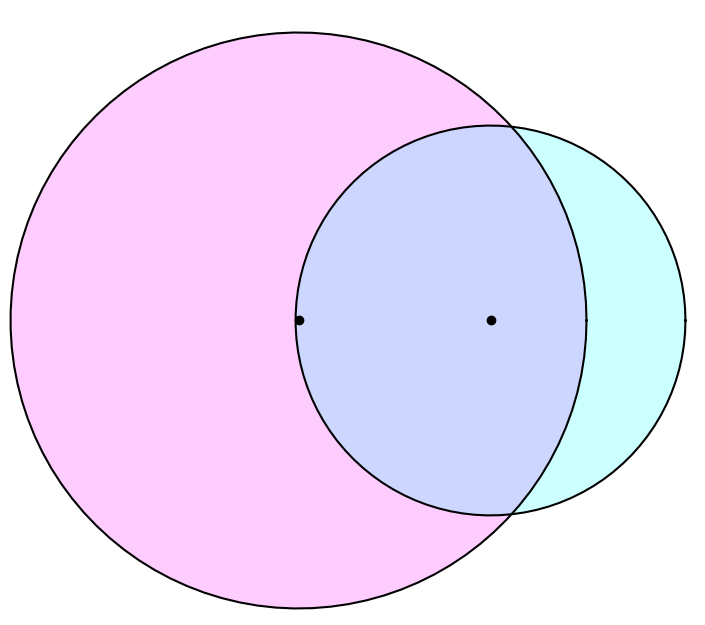}&
\includegraphics[width=4cm]{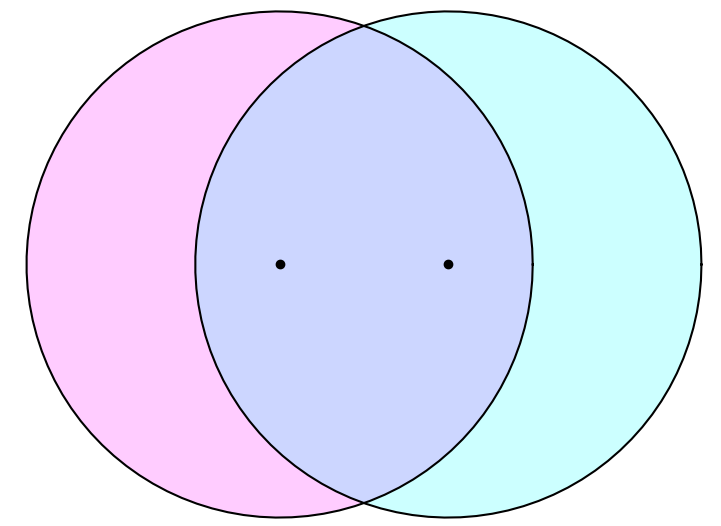}\\
\includegraphics[width=4cm]{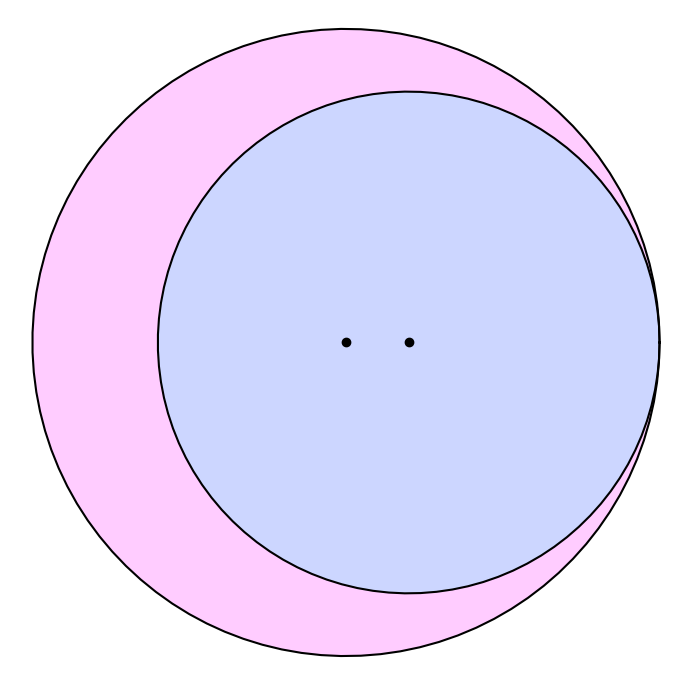}&
\includegraphics[width=4cm]{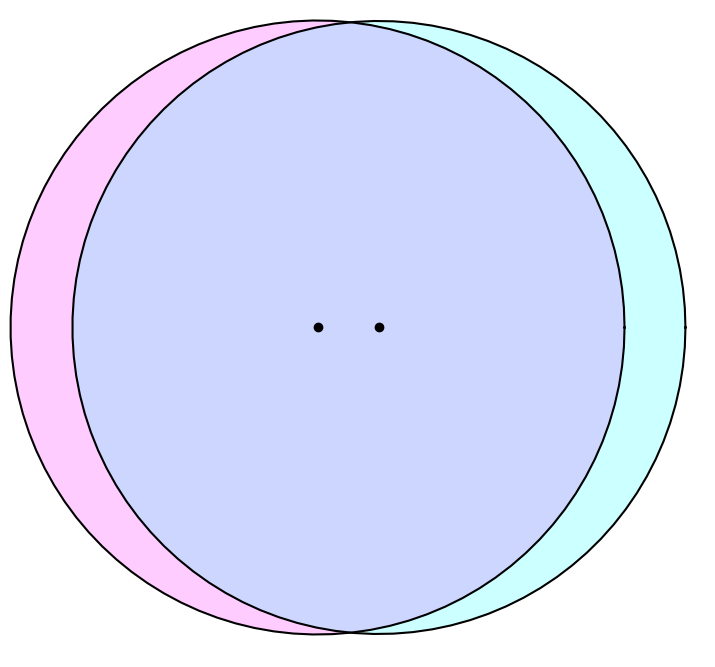}&
\includegraphics[width=4cm]{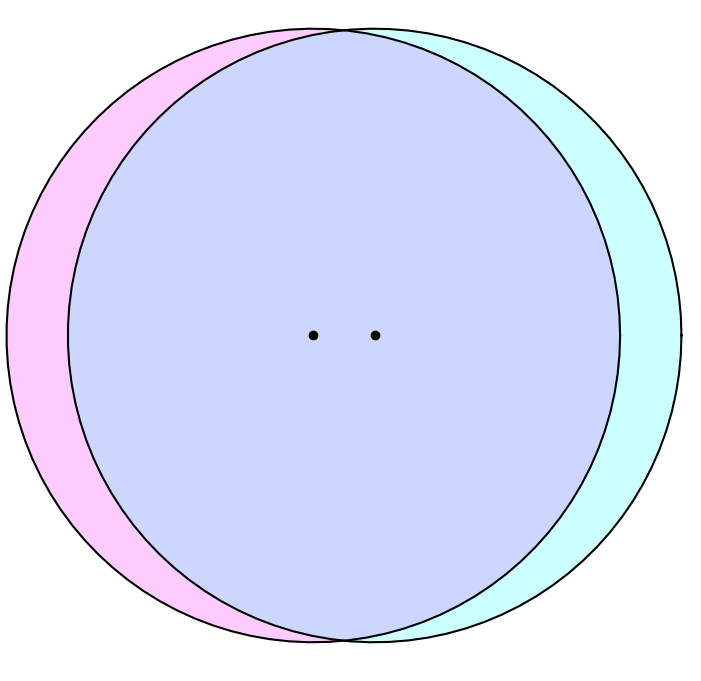}\\
$p = 1$ & $p = 4$ & $p = \infty$\\
\end{tabular}
\caption{Some sets $\CechFt{X,d_{\mu_\Gamma}}{t}$ for $t=1{,}5$ (first row) and $t=5$ (second row).}
\label{Fig_twoballs} 
\end{figure}

Let us show first why one cannot expect a similar result as Proposition \ref{prop_cratio}.
Consider the ambient space $E = \R^2$ and the sets $\Gamma = \{0\}$ and $X = \Gamma \cup \{1\}$. We have $d_{\mu_\Gamma}(1) = 1$ and, for all $t\geq 1$, $\DTMFt{\Gamma}{t}=\overline B(0,t)$ and $\CechFt{X,d_{\mu_\Gamma}}{t}=\overline B(0,t) \cup \overline B\big(1, ( t^p - 1)^\frac{1}{p}\big)$. The sets $\CechFt{X,d_{\mu_\Gamma}}{t}$ are represented in Figure \ref{Fig_twoballs} for $t = 1{,}5$, $t=5$ and several values of $p$.

For $p=1$, the ball $\overline B\big(1, (t^p - 1)^\frac{1}{p}\big)$ is contained in $\overline B(0,t)$. Whereas for $p>1$, the radius $(t^p - 1)^\frac{1}{p}$ is asymptotically equal to $t + o_{t\rightarrow +\infty}(\frac{1}{t^{p-1}})$. Therefore, an $\epsilon \geq 0$ for which the ball $\overline B\big(1, (t^p - 1)^\frac{1}{p}\big)$ would be included in $\overline B(0,t+\epsilon)$ for all $t\geq 0$ should not be lower than $1 = d_H(\Gamma,X)$. Therefore, $d_i(\DTMF{\Gamma}, \CechF{X,d_{\mu_\Gamma}}) = 1 = d_H(\Gamma,X)$.

Even though the filtrations $\DTMF{\Gamma}$ and $\CechF{X,d_{\mu_\Gamma}}$ are distant, the set $\CechFt{X,d_{\mu_\Gamma}}{t}$ is contractible for all $t\geq 0$, and therefore the interleaving distance between the persistence modules $\DTMFpers{\Gamma}$ and $\CechFpers{X,d_{\mu_\Gamma}}$ is 0.

In general, and in the same spirit as Proposition \ref{prop_cratio}, we can obtain a bound on the interleaving distance between the persistence modules $\DTMFpers{\Gamma}$ and $\CechFpers{X,d_{\mu_\Gamma}}$ which does not depend on $X$---see Proposition \ref{prop_DTM_p>1}.

\bigbreak
If $\mu$ is a probability measure on $E$ with compact support $\Gamma$, we define
\[c(\mu, m, p) = \sup_\Gamma (d_{\mu,m}) + \kappa(p)t_\mu(\Gamma), \]
where $\kappa(p) = 1-\frac{1}{p}$, and $t_\mu(\Gamma)$ is the filtration value of the simplex $\Gamma$ in $\CechFnerve{\Gamma, d_\mu}$, the (simplicial) weighted \v{C}ech filtration.
Equivalently, $t_\mu(\Gamma)$ is the value $t$ from which all the balls $\overline B_{d_\mu}(\gamma,t)$, $\gamma \in \Gamma$, share a common point.
\newline \noindent
If $\mu = \mu_{\Gamma}$ is the empirical measure of a finite set $\Gamma \subseteq E$, we denote it $c(\Gamma,m,p)$.
\medbreak

Note that we have the inequality $\frac{1}{2}\mathrm{diam}(\Gamma) \leq t_\mu(\Gamma) \leq 2\mathrm{diam}(\Gamma)$, where $\mathrm{diam}(\Gamma)$ denotes the diameter of $\Gamma$. This follows from writing $t_\mu(\Gamma) = \inf \{t \geq 0, \cap_{\gamma \in \Gamma} \overline B_{d_\mu}(\gamma,t) \neq \emptyset \}$ and using that $\forall \gamma \in \Gamma, d_\mu(\gamma) \leq \mathrm{diam}(\Gamma)$.
%= \inf \{t \geq 0, \cap_{\gamma \in \Gamma} \overline B(\gamma, \big(t^p-f(x)^p\big)^\frac{1}{p}) \}$ 

\begin{proposition}
Let $\mu$ be a measure on $E$ with compact support $\Gamma$, and $d_{\mu}$ be the corresponding DTM of parameter $m$. Consider a set $X \subseteq E$ such that $\Gamma \subseteq X$.
The persistence modules  $\CechFpers{\Gamma, d_{\mu}}$ and $\CechFpers{X, d_{\mu}}$ are $c(\mu, m, p)$-interleaved.

Moreover, if $Y \subseteq E$ is another set such that $\Gamma \subseteq Y$, $\CechFpers{X, d_{\mu}}$ and $\CechFpers{Y, d_{\mu}}$ are $c(\mu, m, p)$-interleaved.

In particular, if $\Gamma$ is a finite set and $\mu = \mu_\Gamma$ its empirical measure, $\DTMFpers{\Gamma}$ and $\CechFpers{X, d_{\mu_\Gamma}}$ are $c(\Gamma, m, p)$-interleaved.
\label{prop_DTM_p>1}
\end{proposition}

The proof involves the two following ingredients, whose proofs are postponed to Subsection \ref{Subsection_proof_prop}.
The first lemma gives a (non-additive) interleaving between the filtrations $\DTMF{\Gamma}$ and $\CechF{X, d_{\mu_\Gamma}}$, relevant for low values of $t$, while the second proposition gives a result for large values of $t$.
  
\begin{lemma}
Let $\mu, \Gamma$ and $X$ be as defined in Proposition \ref{prop_DTM_p>1}. 
Let $\phi\colon t \mapsto 2^{1-\frac{1}{p}} t+ \sup_\Gamma  d_{\mu}$. Then for every $t \geq 0$,
\[ \CechFt{\Gamma, d_{\mu}}{t} \subseteq \CechFt{X, d_{\mu}}{t} \subseteq \CechFt{\Gamma,  d_{\mu}}{\phi(t)}.\] 

\label{lemma_DTMF_mult}
\end{lemma}

In the remainder of the paper, we say that a homology group $H_n(\cdot)$ is trivial if it is of rank 0 when $n>0$, or if it is of rank 1 when $n=0$. We say that a homomorphism between homology groups $H_n(\cdot) \rightarrow H_n(\cdot)$ is trivial if the homomorphism is of rank 0 when $n>0$, or if it is of rank 1 when $n=0$.

\begin{proposition}
Let $\mu, \Gamma$ and $X$ be as defined in Proposition \ref{prop_DTM_p>1}.
Consider the map $v^t_*\colon \CechFperst{X, d_\mu}{t} \rightarrow \CechFperst{X, d_\mu}{t+c}$ induced in homology by the inclusion $v^t\colon \CechFt{X, d_\mu}{t} \rightarrow \CechFt{X, d_\mu}{t+c}$.
If $t \geq t_\mu(\Gamma)$, then $v^t$ is trivial.
\label{prop_stability_DTM}
\end{proposition}

\begin{proof}[Proof of Proposition \ref{prop_DTM_p>1}] 
Denote $c=c(\mu, m, p)$.
For every $t \geq 0$, denote by $v^t\colon \CechFt{X, d_\mu}{t} \rightarrow \CechFt{X, d_\mu}{t+c}$, $w^t\colon \CechFt{\Gamma, d_\mu}{t} \rightarrow \CechFt{\Gamma, d_\mu}{t+c}$ and $j^t\colon \CechFt{\Gamma, d_\mu}{t} \rightarrow \CechFt{X, d_\mu}{t}$ the inclusion maps, and $v^t_*, w^t_*$, and $j^t_*$ the induced maps in homology. 

Notice that, for $t \leq t_\mu(\Gamma)$, the term $2^{1-\frac{1}{p}} t+ \sup_\Gamma  d_{\mu}$ which appears in Lemma \ref{lemma_DTMF_mult} can be bounded as follows: 
\begin{align*}
2^{1-\frac{1}{p}} t + \sup_\Gamma  d_{\mu}
&=  t + (2^{1-\frac{1}{p}}-1) t + \sup_\Gamma  d_{\mu}\\
&\leq t + (2^{1-\frac{1}{p}}-1) t_\mu(\Gamma) +\sup_{\Gamma} d_{\mu_\Gamma} \\ 
&\leq t + (1- \frac{1}{p}) t_\mu(\Gamma) +\sup_{\Gamma} d_{\mu_\Gamma} \\ 
&= t+c
\end{align*}
where, for the second line, we used $2^{1-\frac{1}{p}}-1 \leq 1- \frac{1}{p}$ (Lemma \ref{lemma_ineq_c}).
Consequently, for every $t \leq t_\mu(\Gamma)$, we have $\CechFt{X, d_{\mu}}{t} \subseteq \CechFt{\Gamma, d_{\mu}}{t+c}$.
Thus, for $t \geq 0$, we can define a map $\pi^t\colon \CechFperst{X, d_{\mu}}{t} \rightarrow \CechFperst{\Gamma, d_{\mu}}{t+c}$ as follows: $\pi^t$ is the map induced by the inclusion if $t \leq t_\mu(\Gamma)$, and the zero map if $t \geq t_\mu(\Gamma)$.

The families $(\pi^t)_{t\geq 0}$ and $(j^t_*)_{t\geq 0}$ clearly are $c$-morphisms of persistence modules. Let us show that the pair $((\pi^t)_{t\geq 0}$,$(j^t_*)_{t\geq 0})$ defines a $c$-interleaving between $\CechFpers{\Gamma,d_{\mu}}$ and $\CechFpers{X, d_{\mu}}$. 
\medbreak

Let $t\geq 0$. We shall show that the following diagrams commute:
\begin{center}
\begin{minipage}{0.49\textwidth}
\centering
\begin{tikzcd}
\CechFperst{X, d_\mu}{t} \arrow[dr, swap, "\pi^t"] \arrow[r, "v_*^{t}"] & \CechFperst{X, d_\mu}{t+c} \\
& \CechFperst{\Gamma, d_\mu}{t+c} \arrow[u, swap, "j_*^{t+c}"]  
\end{tikzcd}
\end{minipage}
\begin{minipage}{0.49\textwidth}
\centering
\begin{tikzcd}
\CechFperst{X, d_\mu}{t} \arrow[dr, "\pi^{t}"] & \\
\CechFperst{\Gamma, d_\mu}{t} \arrow[u, "j_*^t"] \arrow[r, swap,"w_*^{t}"] & \CechFperst{\Gamma, d_\mu}{t+c} 
\end{tikzcd}
\end{minipage}
\end{center} 
If $t \leq t_\mu(\Gamma)$, these diagrams can be obtained by applying the homology functor to the inclusions 
\[\CechFt{\Gamma, d_{\mu}}{t} \subseteq \CechFt{X, d_{\mu}}{t} \subseteq \CechFt{\Gamma, d_{\mu}}{t+c} \subseteq \CechFt{X, d_{\mu}}{t+c}.\]
If $t \geq t_\mu(\Gamma)$, the homology group $\CechFperst{\Gamma, d_{\mu}}{t}$ is trivial. Therefore the commutativity of the second diagram is obvious, and the commutativity of the first one follows from Proposition \ref{prop_stability_DTM}.
This shows that $\CechFpers{\Gamma, d_{\mu}}$ and $\CechFpers{X, d_{\mu}}$ are $c$-interleaved.
\medbreak

If $Y$ is another set containing $\Gamma$, define, for all $t \geq 0$, the inclusions $u^t\colon \CechFt{Y, d_\mu}{t} \rightarrow \CechFt{Y, d_\mu}{t+c}$ and $k^t\colon \CechFt{\Gamma, d_\mu}{t} \rightarrow \CechFt{Y, d_\mu}{t+c}$. We can also define a map $\theta^t\colon \CechFperst{Y, d_{\mu}}{t} \rightarrow \CechFperst{\Gamma, d_{\mu}}{t+c}$ as we did for $\pi^t\colon \CechFperst{X, d_{\mu}}{t} \rightarrow \CechFperst{\Gamma, d_{\mu}}{t+c}$.

We can compose the previous diagrams to obtain the following:
\begin{center}
\begin{tikzcd}
\CechFperst{X, d_{\mu}}{t} \arrow[dr, "\pi^{t}"] \arrow[r, "v_*^{t}"] & \CechFperst{X, d_{\mu}}{t+c} \arrow[dr, "\pi^{t+c}"] \arrow[r, "v_*^{t+c}"] & \CechFperst{X, d_{\mu}}{t+2c} \\
& \CechFperst{\Gamma, d_{\mu}}{t+c} \arrow[u, swap, "j_*^{t+c}"] \arrow[d, "k_*^{t+c}"] \arrow[r, swap, "w_*^{t+c}"] & \CechFperst{\Gamma, , d_{\mu}}{t+2c} \arrow[u, swap, "j_*^{t+2c}"] & \\ 
 & \CechFperst{Y, d_{\mu}}{t+c}  \arrow[ur, swap, "\theta^{t+c}"] & 
\end{tikzcd}
\end{center} 
Since all the triangles commute, so does the following:
\begin{center}
\begin{tikzcd}
\CechFperst{X, d_{\mu}}{t} \arrow[dr, swap, "k^{t+c}_*\pi^t"] \arrow[rr, "v^{t+2c}_*"] & & \CechFperst{X, d_{\mu}}{t+2c} \\
& \CechFperst{Y, d_{\mu}}{t+c} \arrow[ur, swap, "j^{t+2c}_* \theta^{t+c}"]  & 
\end{tikzcd}
\end{center} 
We can obtain the same interchanging $X$ and $Y$. Therefore, by definition, the persistence modules $\CechFpers{X, d_{\mu_\Gamma}}$ and $\CechFpers{Y, d_{\mu_\Gamma}}$ are $c$-interleaved, with the interleaving $((k^{t+c}_*\pi^t)_{t\geq 0}, (j^{t+c}_* \theta^{t})_{t\geq 0})$.
\end{proof}

\begin{theorem} \label{thm:DTM-main}
Consider two measures $\mu, \nu$ on $E$ with supports $X$ and $Y$. Let $\mu', \nu'$ be two measures with compact supports $\Gamma$ and $\Omega$ such that $\Gamma \subseteq X$ and $\Omega \subseteq Y$.
We have
\[ d_i(\CechFpers{X,d_\mu},\CechFpers{Y,d_\nu}) \leq 
m^{-\frac{1}{2}}W_2(\mu,\mu') + m^{-\frac{1}{2}}W_2(\mu',\nu') + m^{-\frac{1}{2}}W_2(\nu',\nu) + c(\mu',m,p) + c(\nu',m,p).\]
In particular, if $X$ and $Y$ are finite, we have
\[ d_i(\DTMFpers{X},\DTMFpers{Y}) \leq 
m^{-\frac{1}{2}}W_2(X,\Gamma) + m^{-\frac{1}{2}}W_2(\Gamma,\Omega) + m^{-\frac{1}{2}}W_2(\Omega,Y) + c(\Gamma,m,p) + c(\Omega,m,p).\]
Moreover, with $\Omega = Y$, we obtain
\[ d_i(\DTMFpers{X},\DTMFpers{\Gamma}) \leq 
m^{-\frac{1}{2}}W_2(X,\Gamma) + m^{-\frac{1}{2}}W_2(\Gamma,\Omega) + c(\Gamma,m,p) + c(\Omega,m,p).\]

\label{thm_p>1}
\end{theorem}

\begin{proof}
The proof is the same as Theorem \ref{prop_stability_p=1}, using Proposition \ref{prop_DTM_p>1} instead of Proposition \ref{prop_cratio}.
\end{proof}

Notice that when $p=1$, the constant $c(\Gamma, m, p)$ is equal to the constant $c(\Gamma, m)$ defined in Subsection \ref{subsection_p=1}, and we recover Theorem \ref{prop_stability_p=1} in homology.

\begin{figure}[H]
   \centering
\begin{tabular}{ccc}
\includegraphics[width=4.5cm]{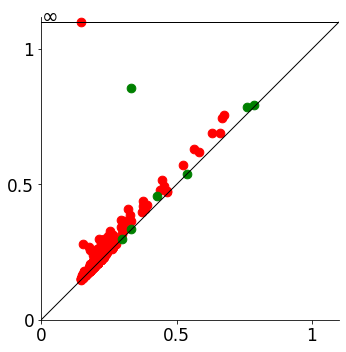}&
\includegraphics[width=4.5cm]{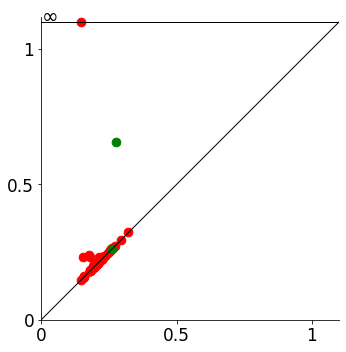}&
\includegraphics[width=4.5cm]{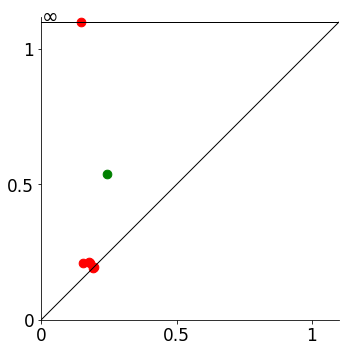}\\
$p=1$ & $p=2$ & 
%$p=3$ & 
%$p=5$ & 
$p=\infty$ \\
\end{tabular}
\caption{Persistence diagrams of the simplicial filtrations $\DTMFrips{X}$ for several values of $p$.}
\label{Fig_diags_p>1}
\end{figure}

\bigbreak
As an illustration of these results, we represent in Figure \ref{Fig_diags_p>1} the persistence diagrams associated to the filtration $\DTMFrips{X}$ for several values of $p$. 
The point cloud $X$ is the one defined in the example page \pageref{example_CechvsDTM}.
Observe that, as stated in Proposition \ref{prop_H0_nonincreasing}, the number of red points (homology in dimension 0) is non-increasing with respect to $p$. 

\subsection{Proof of Lemma \ref{lemma_DTMF_mult} and Proposition \ref{prop_stability_DTM}}
\label{Subsection_proof_prop}
We first prove the lemma stated in the previous subsection.
\begin{proof}[Proof of Lemma \ref{lemma_DTMF_mult}]
Denote $f=d_{\mu}$. Let $x \in X$, and $\gamma$ a projection of $x$ on $\Gamma$. Let us show that for every $t \geq 0$, 
\[\overline B_{f}(x, t) \subseteq \overline B_{f}(\gamma, 2^{1-\frac{1}{p}} t+ f(\gamma)),\]
and the lemma will follow.

Define $d = f(\gamma)$. Let $u \in E$. By definition of the balls, we have 
%$u \in B_f(\gamma, t) \iff t \geq \big(\|u-\gamma\|^p+d^p\big)^{\frac{1}{p}}$ and $	u \in B_f(x, t) \iff t \geq \big(\|u-x\|^p+d_{\mu_\Gamma}(x)^p\big)^{\frac{1}{p}}$.
\[\begin{cases}
	u \in \overline B_{f}(\gamma, t) \iff t \geq \big(\|u-\gamma\|^p+f(\gamma)^p\big)^{\frac{1}{p}}, \\    
	u \in \overline B_{f}(x, t) \iff t \geq \big(\|u-x\|^p+f(x)^p\big)^{\frac{1}{p}}.
\end{cases}\] 
We shall only use 
\[\begin{cases}
	u \in \overline B_{f}(\gamma, t) \impliedby t \geq \|u-\gamma\|+d, \\    
	u \in \overline B_{f}(x, t) \implies t \geq \big(\|u-x\|^p+\|x-\gamma\|^p\big)^{\frac{1}{p}}.
\end{cases}\] 

Let $u \in \overline B_{f}(x,t)$. Let us prove that $u \in \overline B_{f}(\gamma, 2^{1-\frac{1}{p}} t+ d)$.
If $\|u-\gamma\| \leq \|\gamma - x\|$, then $t \geq \|u-\gamma\|$, and we deduce $u \in \overline B_{f}(\gamma, t+d) \subseteq \overline B_{f}(\gamma, 2^{1-\frac{1}{p}}t+d)$.
\smallbreak
Else, we have $\|u-\gamma\| \geq \|\gamma - x\|$. Consider the line segment $[\gamma, u]$ and the sphere $S(\gamma, \|\gamma - x\|)$ of center $\gamma$ and radius $\|\gamma - x\|$. The intersection $S(\gamma, \|\gamma - x\|) \cap [\gamma, u]$ is a singleton. Call its element $x'$. The situation is represented in Figure \ref{Fig_pointx}.
\begin{figure}[H]
\centering
\includegraphics[width=8cm]{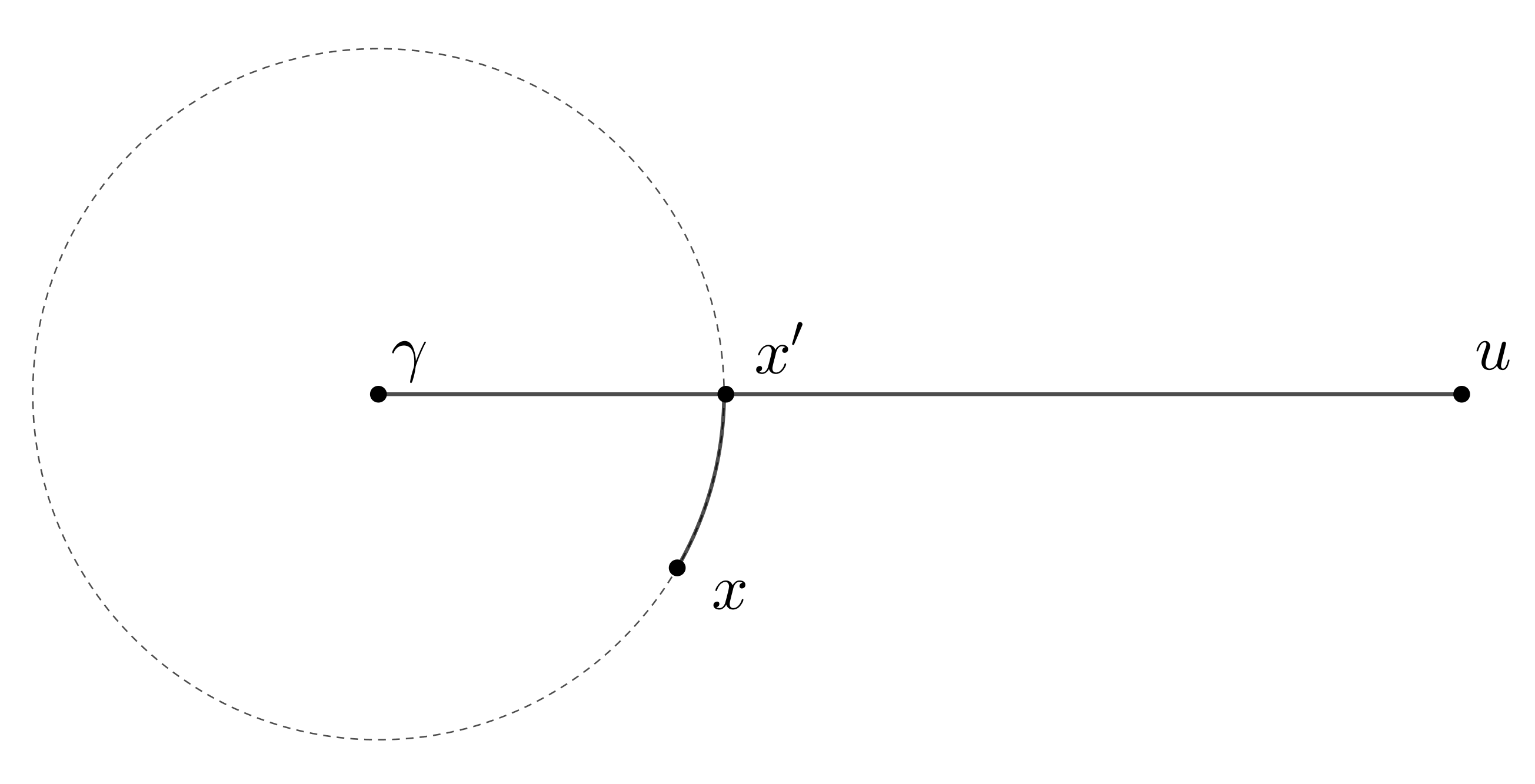}
\caption{Definition of the point $x'$.}
\label{Fig_pointx}
\end{figure}
We have $\|u-x'\|\leq \|u-x\|$ and $\|\gamma-x'\| = \|\gamma-x\|$. 
Therefore 
\[\big(\|u-x'\|^p+\|x'-\gamma\|^p\big)^{\frac{1}{p}} \leq \big(\|u-x\|^p+\|x-\gamma\|^p\big)^{\frac{1}{p}}.\]
We also have $\|\gamma - u\| = \|\gamma - x'\| + \|x' - u\|$ and $\big(\|u-x\|^p+\|x-\gamma\|^p\big)^{\frac{1}{p}} \leq t$. 
Thus it follows from the last inequality that
\[\big(\|u-x'\|^p+(\|u-\gamma\|-\|u-x'\|)^p\big)^{\frac{1}{p}} \leq t.\]
The left-hand term of this inequality is not lower than $2^{\frac{1}{p}-1}\|u - \gamma\|$. Indeed, consider the function 
$s \mapsto \big(s^p+(\|u-\gamma\|-s)^p\big)^{\frac{1}{p}}$ defined for $s \in [0, \|u-\gamma\|]$. 
One shows directly, by computing its derivative, that its minimum is $2^{\frac{1}{p}-1}\|u - \gamma\|$, attained at $s=\frac{\|u - \gamma\|}{2}$. 

We deduce that $2^{\frac{1}{p}-1}\|u - \gamma\| \leq t$, and $\|u - \gamma\| \leq 2^{1-\frac{1}{p}}t$.
Thus $u \in \overline B_{f}(\gamma, 2^{1-\frac{1}{p}} t + d)$.
\end{proof}

Notice that the previous lemma gives a tight bound, as we can see with the following example.
Consider set $\Gamma = \{0\} \subset \R$, $L>0$, and $X = \Gamma \cup \{x\}$ with $x=\frac{L}{2}$. 
Let $m < \frac{1}{2}$, and $f=d_{\mu_\Gamma}$, which is the function distance to $\Gamma$. 
For all $t \geq 2^{\frac{1}{p}-1}L$, we have $L \in \overline B_{f}(x, t)$. Indeed, $r_x(2^{\frac{1}{p}-1}L) = \big((2^{\frac{1}{p}-1}L)^p - (\frac{L}{2})^p\big)^\frac{1}{p} = \frac{L}{2}$.
In comparison, for every $t < \phi(2^{\frac{1}{p}-1}L) = L$, $L \notin \overline B_{f}(0, t)$.

\medbreak
Following this example, we can find a lower bound on the interleaving distance between the persistence modules $\DTMFpers{\Gamma}$ and $\CechFpers{X, d_{\mu_\Gamma}}$. 
Consider $L>0$, the set $\Gamma = \{0,2L\}\subset \R$, $x = \frac{L}{2}$, and $X = \Gamma \cup \{x, 2L-x\}$. 
Let $m < \frac{1}{2}$, and $f=d_{\mu_\Gamma}$. The persistence diagram of the 0th-homology of $\DTMF{\Gamma}$ consists of two points, $(0,+\infty)$ and $(0,L)$.
Regarding $\CechF{X, f}$, the point of finite ordinate in the persistence diagram of its 0th-homology is $(0,2^{\frac{1}{p}-1}L)$. Indeed, for $t = 2^{\frac{1}{p}-1}L$, we have $L \in \overline B_{f}(x, t)$ and $L \in \overline B_{f}(L-x, t)$, hence the set $\CechFt{X, d_{\mu_\Gamma}}{t}$ is connected.
We deduce that these persistence modules are at least $(1-2^{\frac{1}{p}-1})L$-interleaved.

In comparison, the upper bound we prove in Proposition \ref{prop_DTM_p>1} is $(1-\frac{1}{p})L$.

\bigbreak
We now prove the proposition stated in the previous subsection.

\begin{proof}[Proof of Proposition \ref{prop_stability_DTM}]
Denote $f=d_{\mu}$. Let $t \geq t_\mu(\Gamma)$. By definition of $t_\mu(\Gamma)$, there exists a point $O_\Gamma \in \bigcap_{\gamma \in \Gamma} \overline B_f(\gamma, t_\mu(\Gamma))$. 

In order to show that $v^t_*\colon \CechFperst{X, d_\mu}{t} \rightarrow \CechFperst{X, d_\mu}{t+c}$ is trivial, we introduce an intermediate set between $\CechFt{X, d_{\mu_\Gamma}}{t}$ and $\CechFt{X, d_{\mu_\Gamma}}{t+c}$:
\[\begin{cases}
\CechFt{X, d_{\mu_\Gamma}}{t} &= \bigcup_{x \in X \setminus \Gamma} \overline B_f(x,t) \cup \bigcup_{\gamma \in \Gamma} \overline B_f(\gamma, t), \\    
\widetilde V^t &:= \bigcup_{x \in X \setminus \Gamma} \overline B_f(x,t) \cup \bigcup_{\gamma \in \Gamma} \overline B_f(\gamma, t+c),\\
\CechFt{X, d_{\mu_\Gamma}}{t+c} &= \bigcup_{x \in X \setminus \Gamma} \overline B_f(x,t+c) \cup \bigcup_{\gamma \in \Gamma} \overline B_f(\gamma, t+c).
\end{cases}\] 
Since $t \geq t_\mu(\Gamma)$, we have $O_\Gamma \in \widetilde V^t$. Let us show that $\widetilde V^t$ is star-shaped around $O_\Gamma$. 

Let $x \in X$ and consider $\gamma$ a projection of $x$ on $\Gamma$. We first prove that $\overline B_f(x,t) \cup \overline B_f(\gamma, t+c)$ is star-shaped around $O_\Gamma$. 
Let $y \in \overline B_f(x,t)$. We have to show that the line segment $[y, O_\Gamma]$ is a subset of $\overline B_f(x,t) \cup \overline B_f(\gamma, t+c)$. 
Let $D$ be the affine line going through $y$ and $O_\Gamma$, and denote by $q$ the orthogonal projection on $D$. We have $[y, O_\Gamma] \subseteq [y, q(x)] \cup [q(x), O_\Gamma]$. The first line segment $[y, q(x)]$ is a subset of $\overline B_f(x,t)$. 
Regarding the second line segment $[q(x), O_\Gamma]$, let us show that $q(x) \in \overline B_f(\gamma, t+c)$, and $[q(x), O_\Gamma] \subseteq \overline B_f(\gamma, t+c)$ will follow. 
The situation is pictured in Figure \ref{Fig_proof_p>1}.

\begin{figure}[H]
\centering
\begin{tabular}{cc}
\includegraphics[width=7.0cm]{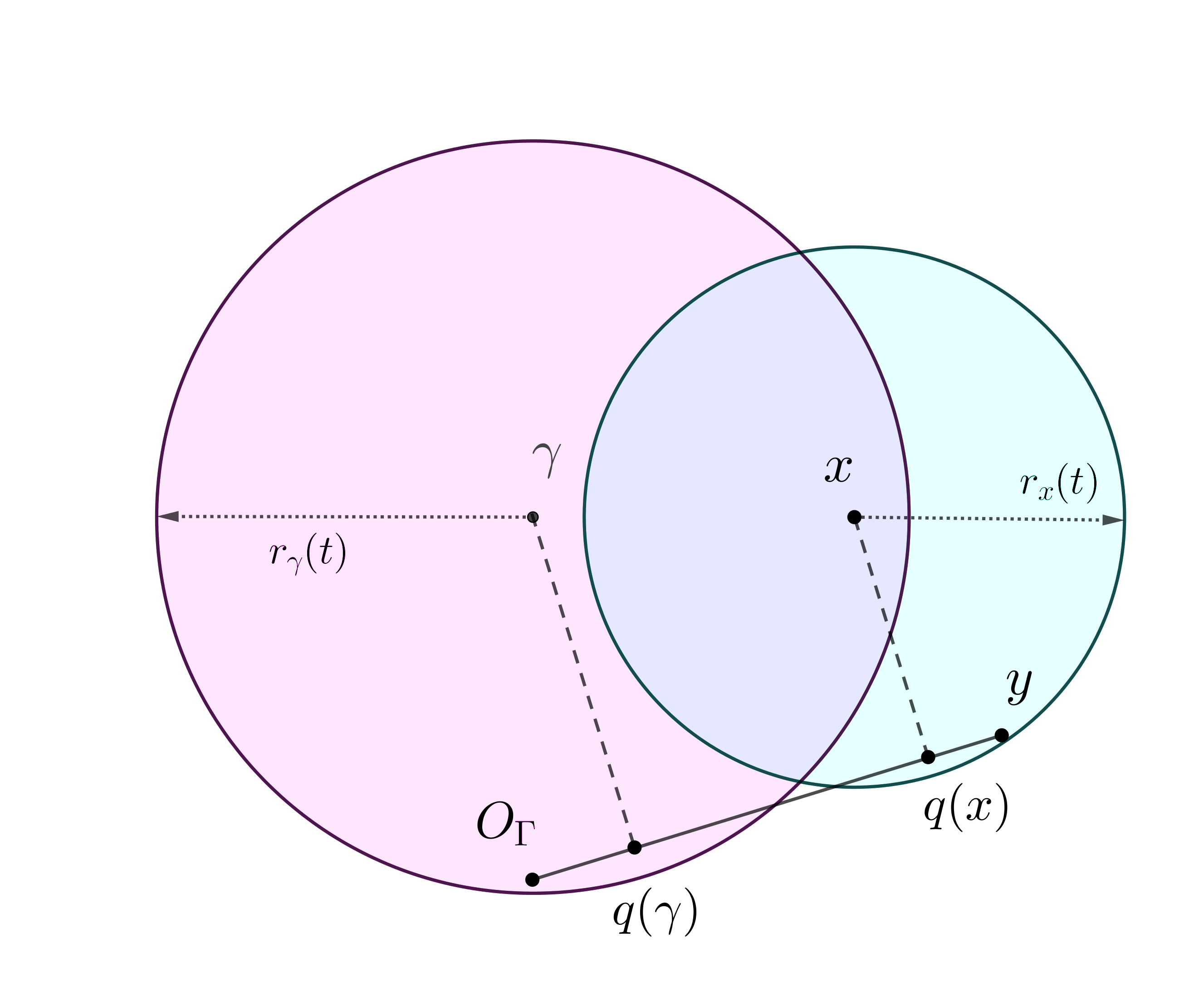} &
\includegraphics[width=7.0cm]{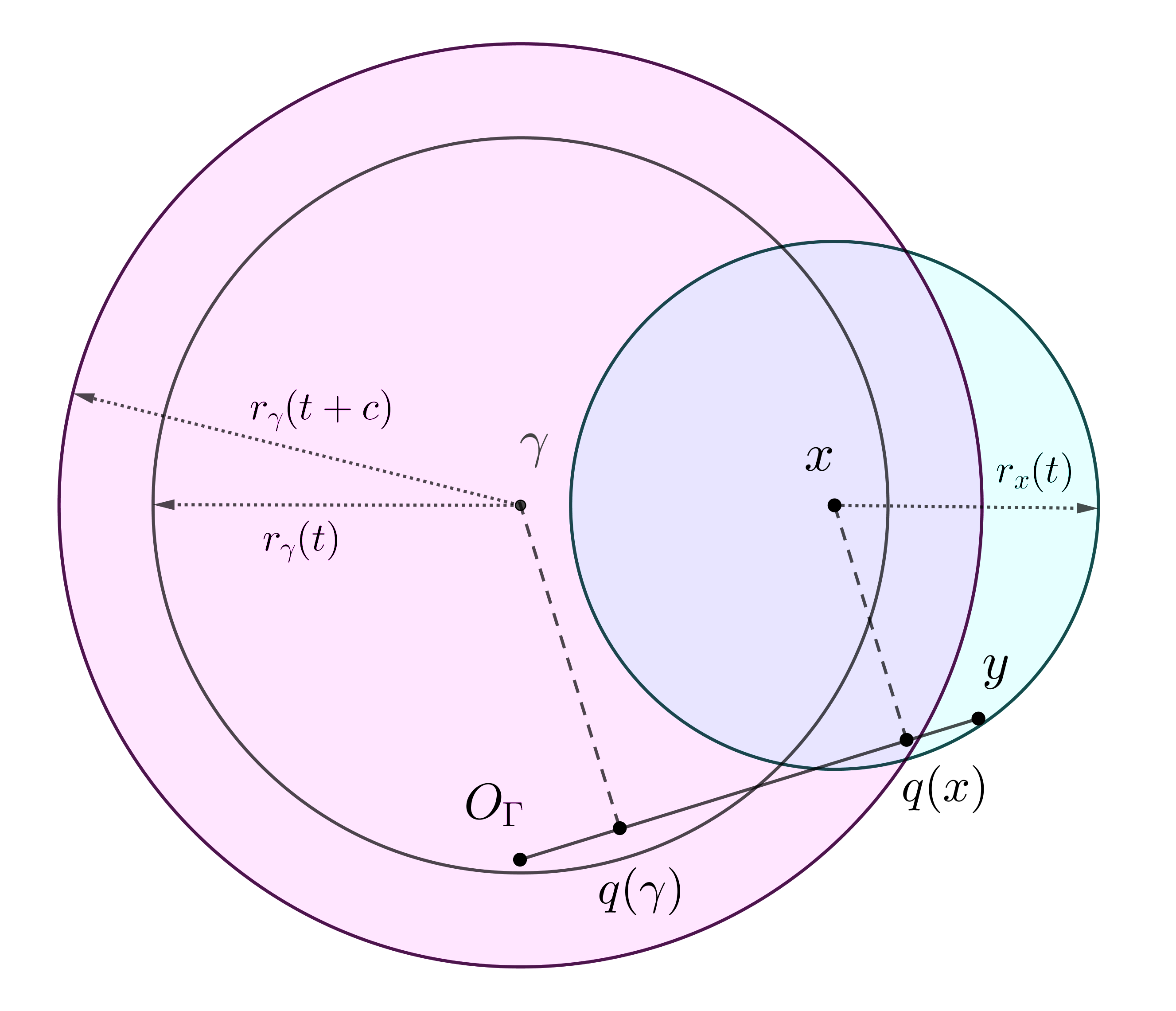} \\
$\overline B_f(x,t) \cup \overline B_f(\gamma, t)$ &
$\overline B_f(x,t) \cup \overline B_f(\gamma, t+c)$ \\
\end{tabular}
\caption{Construction of an intermediate set $\widetilde V^t$.}
\label{Fig_proof_p>1}
\end{figure}
\noindent
%Denote by $q(\gamma)$ the orthogonal projection of $\gamma$ on $D$. 
According to Lemma \ref{lemma_right_trapezoid},  
\[ \|\gamma - q(x)\|^2 \leq \|x-\gamma\|^2 + \|x-q(x)\|(2\|\gamma-q(\gamma)\|-\|x-q(x)\|). \]
Let $d = \|x-q(x)\|$.
Since $d=\|x - q(x)\| \leq  \big(t^p-d_{\mu}(x)^p\big)^{\frac{1}{p}} \leq \big(t^p-\|x-\gamma\|^p\big)^{\frac{1}{p}}$, we have $\|x-\gamma\| \leq (t^p - d^p)^\frac{1}{p}$.
Moreover, $\|\gamma-q(\gamma)\| \leq \|\gamma-O_\Gamma\| \leq t_\mu(\Gamma)$.
The last inequality then gives
\[ \|\gamma - q(x)\|^2 \leq (t^p - d^p)^\frac{2}{p} + d(2t_\mu(\Gamma)-d). \]
According to Lemma \ref{lemma_inequality_DTM}, we obtain that $\|\gamma - q(x)\|$ is not greater than $t+\kappa(p) t_\mu(\Gamma)$. 
Therefore, we have the inequality 
\[\big( (t+\kappa(p) t_\mu(\Gamma) + f(\gamma))^p-f(\gamma)^p\big)^\frac{1}{p} \geq \big(t+\kappa(p) t_\mu(\Gamma) + f(\gamma)\big)-f(\gamma) \geq \|\gamma - q(x)\|,\] 
and we deduce 
$q(x) \in \overline B_f\big(\gamma, t+\kappa(p) t_\mu(\Gamma) + f(\gamma)\big) \subset \overline B_f(\gamma, t+c)$.

In conclusion, $[y, O_\Gamma] \subset \overline B_f(x,t) \cup \overline B_f(\gamma, t+c)$. This being true for every $y \in \overline B_f(x,t)$, and obviously true for $y \in \overline B_f(\gamma,t+c)$, we deduce that $\overline B_f(x,t) \cup \overline B_f(\gamma, t+c)$ is star-shaped around $O_\Gamma$. 
Finally, since $O_\Gamma \in \bigcap_{\gamma \in \Gamma} \overline B_f(\gamma, t_X(\Gamma))$, we have that $\widetilde V^t$ is star-shaped around $O_\Gamma$. 

\medbreak
To conclude, notice that the map $v_*^t$ factorizes through $H_*(\widetilde V^t)$.
Indeed, consider the diagram of inclusions: 
\begin{center}
\begin{tikzcd}
\CechFt{X, d_{\mu_\Gamma}}{t} \arrow[""]{r}  \arrow[bend right=-20, "v^t"]{rr} & \widetilde V^t \arrow[""]{r} & \CechFt{X, d_{\mu_\Gamma}}{t+c}
.\end{tikzcd}
\end{center}
Applying the singular homology functor, we obtain
\begin{center}
\begin{tikzcd}
\CechFperst{X, d_{\mu_\Gamma}}{t} \arrow[""]{r}  \arrow[bend right=-20, "v^t_*"]{rr} & H_*(\widetilde V^t) \arrow[""]{r} & \CechFperst{X, d_{\mu_\Gamma}}{t+c}
.\end{tikzcd}
\end{center}
Since $\widetilde V^t$ is star-shaped, $H_*(\widetilde V^t)$ is trivial, and so is $v^t_*$.
\end{proof}

\section{Conclusion}
In this paper we have introduced the DTM-filtrations that depend on a parameter $p\geq 1$. This new family of filtrations extends the filtration introduced in \cite{Buchet_Efficient} that corresponds to the case $p=2$. 

The established stability properties are, as far as we know, of a new type: the closeness of two DTM-filtrations associated to two data sets relies on the existence of a well-sampled underlying object that approximates both data sets in the Wasserstein metric. This makes the DTM filtrations robust to outliers. 
Even though large values of $p$ lead to persistence diagrams with less points in the 0th homology, the choice of $p=1$ gives the strongest stability results.
When $p>1$, the interleaving bound is less significant since it involves the diameter of the underlying object, but the obtained bound is consistent with the case $p=1$ as it converges to the bound for $p=1$ as $p$ goes to $1$.

It is interesting to notice that the proofs rely on only a few properties of the DTM. As a consequence, the results should extend to other weight functions, such that the DTM with an exponent parameter different from 2, or kernel density estimators. Some variants concerning the radius functions in the weighted \v{C}ech filtration, are also worth considering. 
%The analysis shows that their asymptotic behaviour should look like the one of the case $p=1$.
The analysis shows that one should choose radius functions whose asymptotic behaviour look like the one of the case $p=1$.
In the same spirit as in \cite{sheehy13linear, Buchet_Efficient} where sparse-weighted Rips filtrations were considered, it would also be interesting to consider sparse versions of the DTM-filtrations and to study their stability properties. 

Last, the obtained stability results, depending on the choice of underlying sets, open the way to the statistical analysis of the persistence diagrams of the DTM-filtrations, a problem that will be addressed in a further work. 

\appendix

\section{Supplementary results for Section \ref{section_Vp_modules}}
\begin{lemma}
\label{lemma_inequality_stability}
Let $c, \epsilon$ and $x$ be non-negative real numbers, and $t \geq a$. Define $\alpha = (1+c^p)^\frac{1}{p}$ and $k = \epsilon \alpha$. Then $t+ k \geq a + c \epsilon$, and
\[\big( (t + k)^p-(x+c \epsilon)^p \big)^\frac{1}{p}-(t^p-x^p)^\frac{1}{p} \geq \epsilon\]
\end{lemma}

\begin{proof}
Let $\mathcal{D} = \{(t,x), t \geq x \geq 0 \} \subseteq \R^2$. Let us find the minimum of 
\[ \begin{array}{ccccl}
\Phi &: &\mathcal{D} &\longrightarrow &\R \\
& &(t,x) &\longmapsto &\big( (t+\alpha \epsilon)^p-(x+c\epsilon)^p \big)^\frac{1}{p}-(t^p-x^p)^\frac{1}{p}
\end{array} \]
%Its interior is denoted by  $\mathring{\mathcal{D}} = \{(t,x), t > x > 0 \}$.
An $x > 0$ being fixed, we study $\phi\colon t \mapsto \Phi(t, x)$ on the interval $(x, +\infty)$. Its derivative is
\[\phi'(t) = \frac{(t+\alpha \epsilon)^{p-1}}{\big((t+\alpha \epsilon)^p - (x + c\epsilon)^p\big)^{1-\frac{1}{p}}}-\frac{t^{p-1}}{(t^p - x^p)^{1-\frac{1}{p}}}\]
We solve:
\begin{align*}
\phi'(t) = 0 &\iff (t+\alpha \epsilon)^{p-1}(t^p - x^p)^{1-\frac{1}{p}} = t^{p-1}((t+\alpha \epsilon)^p - (x + c\epsilon)^p)^{1-\frac{1}{p}} \\
&\iff \underbracket[.8pt]{(t+\alpha \epsilon)^{p}(t^p} - x^p)= \underbracket[.8pt]{t^{p}((t+\alpha \epsilon)^p} - (x + c\epsilon)^p) \\
&\iff (t+\alpha \epsilon)^{p}x^p= t^{p}(x + c\epsilon)^p \\
&\iff \frac{t+\alpha \epsilon}{t} = \frac{x+c \epsilon}{x} \\
&\iff t = \frac{\alpha}{c} x 
\end{align*}
We obtain the second line by raising the equality to the power of $\frac{p}{p-1}$.
Hence the derivative of $\phi$ vanishes only at $t = \frac{\alpha}{c} x$. Together with $\lim_{+\infty} \phi = +\infty$, we deduce that $\phi$ attains its minimum at $t = x$ or $t = \frac{\alpha}{c} x$.

Let us show that $\phi(\frac{\alpha}{c} x) = \epsilon$.
\[\begin{array}{cclcl}
\phi(\frac{\alpha}{c} x) = \Phi(\frac{\alpha}{c} x, x) 
&= &\big((\frac{\alpha}{c} x+\alpha \epsilon)^p-(x+c\epsilon)^p\big)^\frac{1}{p} &- &\big((\frac{\alpha}{c} x)^p-x^p\big)^\frac{1}{p} \\
&= &\big((\frac{\alpha}{c})^p (x+ c\epsilon)^p-(x+c\epsilon)^p\big)^\frac{1}{p} &- &x\big((\frac{\alpha}{c} )^p-1\big)^\frac{1}{p} \\
&= &(x+c\epsilon)\big((\frac{\alpha}{c})^p-1\big)^\frac{1}{p} &- &x\big((\frac{\alpha}{c} )^p-1\big)^\frac{1}{p} \\
&=& \multicolumn{3}{l}{c\epsilon \big( (\frac{\alpha}{c})^p-1 \big)^\frac{1}{p}}
\end{array}\]
Using $\alpha = (1+c^p)^\frac{1}{p}$, one obtains that $c \big( (\frac{\alpha}{c})^p-1 \big)^\frac{1}{p} = 1$. Therefore, $\phi(\frac{\alpha}{c} x) = \epsilon$.

Secondly, consider $\Phi$ on the interval $\{(x,x), x\geq 0\}$. 
\newline \noindent
The function $t \mapsto \Phi(x,x) = ((x+\alpha \epsilon)^p-(x+c\epsilon)^p)^\frac{1}{p}$ is increasing. Its minimum is $\Phi(0,0) = ((\alpha \epsilon)^p-(c\epsilon)^p)^\frac{1}{p}=\epsilon(\alpha^p-c^p)^\frac{1}{p} = \epsilon$.

In conclusion, on every interval $(x, +\infty) \times \{x\} \subseteq \mathcal{D}$, $\Phi$ admits $\epsilon$ as a minimum. Therefore, $\epsilon$ is the minimum of $\Phi$ on $\mathcal{D}$.
\end{proof}

\section{Supplementary results for Section \ref{section_DTM}}
\begin{lemma}
For all $p \geq 1$, $2^{1-\frac{1}{p}}-1 \leq 1 - \frac{1}{p}$.
\label{lemma_ineq_c}
\end{lemma}

\begin{proof}
The convexity property of the function $x \mapsto 2^x$ gives, for all $x \in [0,1]$, $2^x \leq x+1$.  
Hence $2^{1-\frac{1}{p}}-1 \leq 1 - \frac{1}{p}$.
\end{proof}

\begin{lemma}
Let $\gamma, x \in E$, $D$ an affine line, and $q(\gamma), q(x)$ the projections of $\gamma$ and $x$ on $D$. Then
\[ \|\gamma - q(x)\|^2 \leq \|x-\gamma\|^2 + \|x-q(x)\|(2\|\gamma-q(\gamma)\|-\|x-q(x)\|). \]

\label{lemma_right_trapezoid}
\end{lemma}

\begin{proof}
We first study the case where $\gamma, x$ and $D$ lie in the same affine plane. If $\gamma$ and $x$ are on opposite sides of $D$, the result is obvious. Otherwise, the points $\gamma, x, q(\gamma)$ and $q(x)$ form a right trapezoid (see Figure \ref{Fig_trapezoid}).

\begin{figure}[H]
\centering
\includegraphics[width=6.5cm]{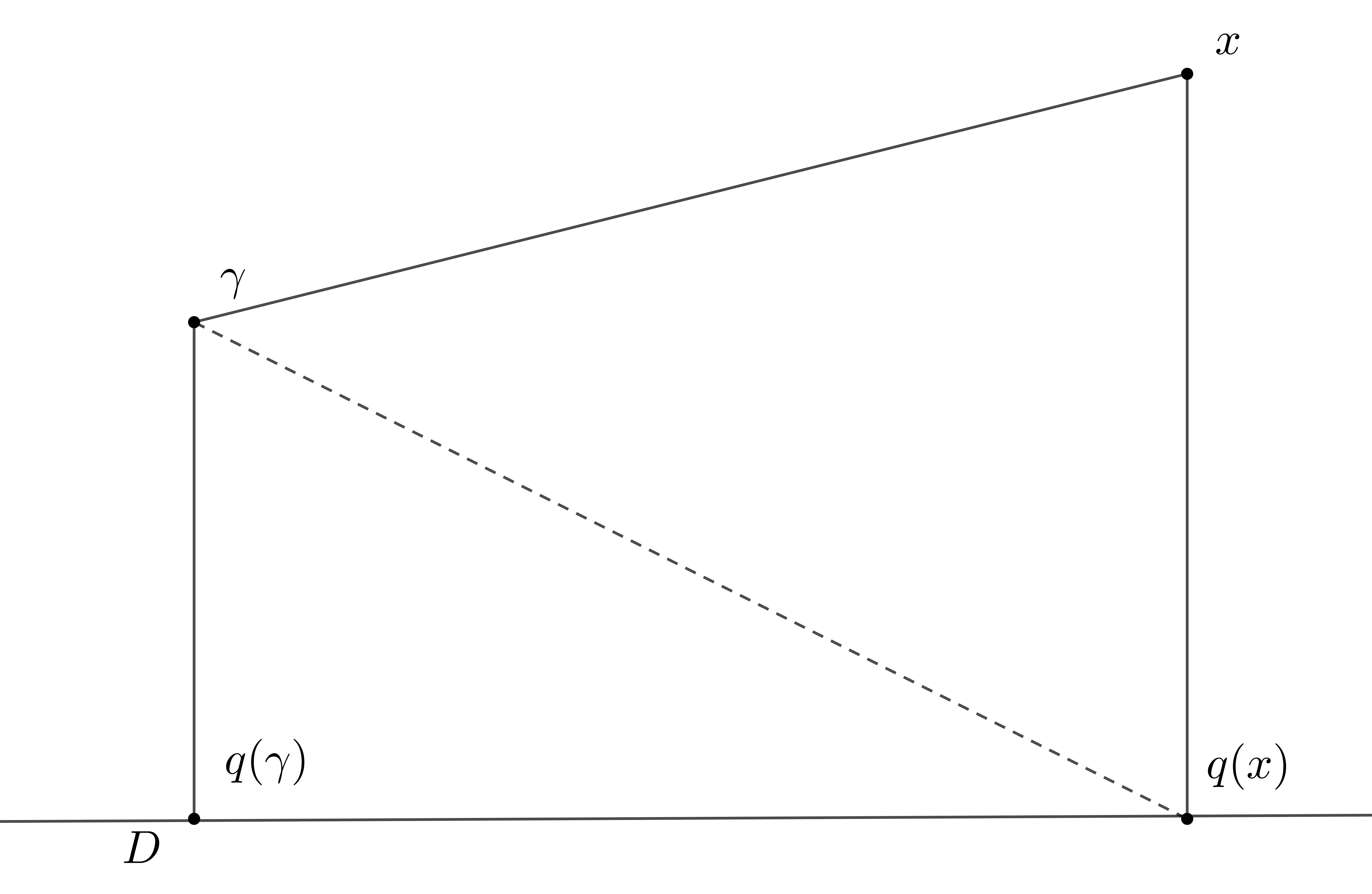}
\caption{The points $\gamma, x, q(\gamma)$ and $q(x)$ form a right trapezoid.}
\label{Fig_trapezoid}
\end{figure}

Using the Pythagorean theorem on the orthogonal vectors $\gamma - q(\gamma)$ and $q(\gamma) - q(x)$, and on $(\gamma - q(\gamma)) - (x - q(x))$ and $q(\gamma)-q(x)$, we obtain
\[\begin{cases}
	\|\gamma - q(\gamma)\|^2 +  \|q(\gamma) - q(x)\|^2 = \|\gamma - q(x)\|^2, \\    
	\|(\gamma - q(\gamma)) - (x - q(x))\|^2 + \|q(\gamma)-q(x)\|^2 = \|\gamma-x\|^2.
\end{cases}\] 
Using that $\|(\gamma - q(\gamma)) - (x - q(x))\| = \mid \|\gamma - q(\gamma)\| - \|x - q(x)\| \mid$, the second equality rephrases as $\|q(\gamma)-q(x)\|^2 = \|\gamma-x\|^2 -  (\|\gamma - q(\gamma)\| - \|x - q(x)\|)^2$.
Combining these two equalities gives 
\begin{align*}
\|\gamma - q(x)\|^2 &= \|\gamma - q(\gamma)\|^2 +  \|q(\gamma) - q(x)\|^2 \\
&= \|\gamma - q(\gamma)\|^2 + \|\gamma-x\|^2 -  (\|\gamma - q(\gamma)\| - \|x - q(x)\|)^2\\
&= \|\gamma-x\|^2 + \|x - q(x)\|(2\|\gamma - q(\gamma)\|-\|x - q(x)\|). 
\end{align*}

Now, if $\gamma, x$ and $D$ do not lie in the same affine plane, denote by $P$ the affine plane containing $D$ and $x$. Let $\tilde \gamma$ the point of $P$ such that $\|\gamma-q(\gamma)\| = \|\tilde \gamma-q(\gamma)\|$ and $\|\gamma-q(x)\| = \|\tilde \gamma-q(x)\|$. Using the previous result on $\tilde \gamma$ and the inequality $\|\gamma-x\| \geq \|\tilde \gamma-x\|$, we obtain the result.
\end{proof}

\label{proof_inequality_DTM}
\begin{lemma}
%If $t(\sigma) \geq t(\tau)$ For all $d \in []$
%\[ (t(X)^p - d^p)^\frac{2}{p} + d(2t(\Gamma)-d) \leq (t(X)+\kappa t(\Gamma))^2    \] 
Let $a,b,d \geq 0$ such that $b \leq a$ and $d \leq a$. Then 
%$0\leq d \leq a$, $b\leq a$
\[ (a^p - d^p)^\frac{2}{p} + d(2b-d) \leq (a+\kappa b)^2,\]
with $\kappa = 1-\frac{1}{p}$. 
%with $\kappa = 2^{1-\frac{1}{p}}-1$. 
\label{lemma_inequality_DTM}
\end{lemma}
\begin{proof}
The equation being homogeneous with respect to $a$, it is enough to show that
\[(1 - d^p)^\frac{2}{p} + d(2b-d) \leq (1+\kappa b)^2\]
 with $b\leq 1$ and $d \leq 1$.
We shall actually show that $(1 - d^p)^\frac{2}{p} + d(2b-d) \leq 1+2 \kappa b$.

Note that this is true when $d \leq \kappa$. Indeed, $(1 - d^p)^\frac{2}{p} + d(2b-d) \leq 1  + 2db \leq 1  + 2\kappa b$. 
%When $b \leq \kappa$, the inequality is also clearly true, since $(1 - d^p)^\frac{2}{p} + d(2b-d) \leq 1 + \frac{1}{4}b^2 \leq 1 + \frac{1}{4}b\kappa$.
Now, notice that it is enough to show the inequality for $b=1$. Indeed, it is equivalent to $(1 - d^p)^\frac{2}{p} - 1 - d^2 \leq 2\kappa b - 2 d b = 2b(\kappa - d)$. For every $d \geq \kappa$, the right-hand side of this inequality is nonpositive, hence the worst case happens when $b=1$. What is left to show is the following: $\forall d \in [\kappa, 1]$,
\[(1 - d^p)^\frac{2}{p} + d(2-d) \leq 1  + 2\kappa. \]

The function $x \mapsto (1-x)^\frac{1}{p}$ being concave on $[0,1]$, we have $(1-x)^\frac{1}{p} \leq 1 - \frac{1}{p}x$ for all $x \in [0,1]$. Therefore, $(1 - d^p)^\frac{1}{p} \leq 1 - \frac{1}{p}d^p$.
Consider the function 
\[\phi\colon d \mapsto (1 - \frac{1}{p}d^p)^2 + d(2-d).\] 
Let us show that $\forall d \in [0,1], \phi(d) \leq 1+2 \kappa$.

%Notice that $d \mapsto (1 - \frac{1}{p}d^p)^2$ is decreasing on $[0,1]$, and $d \mapsto d(2b-d)$ is decreasing on $[b,1]$. Therefore, $\phi$ admits a maximum in $[0,1]$, hence it is enough to show that $\forall d \in [0,b], \phi(d) \leq 1+2 \kappa b$.

This inequality is obvious for $d = 0$. It is also the case for $d=1$, since we obtain $(1 - \frac{1}{p}d^p)^2 + d(2-d) = (1 - \frac{1}{p})^2 + 1 = \kappa^2 +1 $. 
On the interval $[0,1]$, the derivative of $\phi$ is $\phi'(d) = \frac{2}{p}d^{2p-1}-2d^{p-1}-2d+2$.
Let $d_*$ be such that $\phi'(d_*) = 0$. 
Multiplying $\phi'(d_*)$ by $\frac{d_*}{2}$ gives the relation $\frac{1}{p}d_*^{2p}-d_*^{p}-d_*^2+d_*=0$.
Subtracting this equality in $\phi(d_*)$ gives $\phi(d_*) = 1 -(\frac{1}{p}-\frac{1}{p^2})d_*^{2p}+(1-\frac{2}{p})d_*^p + d_*$. We shall show that the following function $\psi$, defined for all $d \in [0,1]$, is not greater than $1+2 \kappa$:
\[\psi\colon d \longmapsto 1 - \frac{1}{p}(1-\frac{1}{p})d^{2p}+(1-\frac{2}{p})d^p + d.\]

We consider the cases $p\geq 2$ and $p \leq 2$ separately. 
In each case, $1-\frac{1}{p}\geq 0$.
Assume that $p\geq 2$. Then $d^p \leq 1$ and $1-\frac{2}{p} \geq 0$. Therefore $(1-\frac{2}{p})d^p \leq 1-\frac{2}{p}$, and we obtain
\begin{align*}
\psi(d) &\leq 1 + (1-\frac{2}{p})d^p + d \\
&\leq 1+ (1-\frac{2}{p}) d + d\\
& = 1+ 2(1-\frac{1}{p})
\end{align*}

Now assume that $p\leq 2$.
We have the following inequality: $d - d^p \leq p-1$. Indeed, by considering its derivative, one shows that the application $d \mapsto d - d^p$ is maximum for $d = p^{-\frac{1}{p-1}}$, for which
\begin{align*}
d - d^p = d(1 - d^{p-1}) &= p^{-\frac{1}{p-1}}(1- p^{-1}) \\
&= p^{-\frac{1}{p-1}-1}(p- 1) \\
&= p^{-\frac{p}{p-1}}(p- 1) \leq p - 1. 
\end{align*}
Using $(\frac{2}{p}-1) \geq 0$ and $d^p \geq d - (p-1)$, we obtain $(\frac{2}{p}-1)d^p \geq (\frac{2}{p}-1)d - (\frac{2}{p}-1)(p-1)$.
Going back to $\psi(d)$, we have
\begin{align*}
\psi(d) &= 1 - \frac{1}{p}(1-\frac{1}{p})d^{2p}-(\frac{2}{p}-1)d^p + d \\
&\leq 1 - \frac{1}{p}(1-\frac{1}{p})d^{2p}-(\frac{2}{p}-1)d + (\frac{2}{p}-1)(p-1) + d \\
&=  1 - \frac{1}{p}(1-\frac{1}{p})d^{2p}+(2-\frac{2}{p})d + (\frac{2}{p}-1)(p-1).
\end{align*}
Let us verify that $d \mapsto 1 - \frac{1}{p}(1-\frac{1}{p})d^{2p}+2(1-\frac{1}{p})d + (\frac{2}{p}-1)(p-1)$ is increasing. Its derivative is
\begin{align*}
-2p\frac{1}{p}(1-\frac{1}{p})d^{2p-1}+2(1-\frac{1}{p}) &\geq -2p\frac{1}{p}(1-\frac{1}{p})+2(1-\frac{1}{p}) \\
&= 0
\end{align*}
We deduce that $\psi(d) \leq \psi(1)$ for all $d \in [0,1]$. 
The value $\psi(1)$ is $1 - \frac{1}{p}(1-\frac{1}{p})+2(1-\frac{1}{p}) + (\frac{2}{p}-1)(p-1)$. 
Moreover, we have $-\frac{1}{p}(1-\frac{1}{p})+ (\frac{2}{p}-1)(p-1) \leq 0$. Indeed, $-\frac{1}{p}(1-\frac{1}{p})+ (\frac{2}{p}-1)(p-1) = -\frac{(p-1)^3}{p^2}$.
Therefore $\psi(1) \leq 1 + 2(1-\frac{1}{p})$.
\end{proof}

\bibliographystyle{plain} %ordre alphabetique
\bibliography{bib_Filtrations}

\end{document}